# Highly Efficient siRNA Delivery from Core-Shell Mesoporous Silica Nanoparticles with Multifunctional Polymer Caps

Karin Möller[1], Katharina Müller[2], Hanna Engelke[1], Christoph Bräuchle[1], Ernst Wagner*,[2] and Thomas Bein*,[1]

[1]Department of Chemistry and Center for NanoScience, University of Munich (LMU), Butenandtstrasse 5–13, 81377 Munich, Germany
Fax: (+49) 89-2180-77622

[2]Pharmaceutical Biotechnology and Center for NanoScience, University of Munich (LMU), Butenandtstrasse 5-13, 81377 Munich, Germany

Emails: ernst.wagner@cup.uni-muenchen.de, bein@lmu.de

**Abstract**

A new general route for siRNA delivery is presented combining porous core-shell silica nanocarriers with a modularly designed multifunctional block copolymer. Specifically, the internal storage and release of siRNA from mesoporous silica nanoparticles (MSN) with orthogonal core-shell surface chemistry was investigated as a function of pore-size, pore morphology, surface properties and pH. Very high siRNA loading capacities of up to 380 µg/mg MSN were obtained with charge-matched amino-functionalized mesoporous cores, and release profiles show up to 80% siRNA elution after 24 h. We demonstrate that adsorption and desorption of siRNA is mainly driven by electrostatic interactions, which allow for high loading capacities even in medium-sized mesopores with pore diameters down to 4 nm in a stellate pore morphology. The negatively charged MSN shell enabled the association with a block copolymer containing positively charged artificial amino acids and oleic acid blocks, which acts simultaneously as capping function and endosomal release agent. The potential of this multifunctional delivery platform is demonstrated by highly effective cell transfection and siRNA delivery into KB-cells. A luciferase reporter gene knock-down of up to 90% was possible using extremely low cell exposures with only 2.5 µg MSN containing 32 pM siRNA per 100 µL well.

**Introduction**

In the past decade mesoporous silica nanoparticles (MSN) have been recognized as a powerful general tool for packaging fragile or toxic pharmaceuticals and ensuring their safe passage through cell membranes, to finally unload their cargo at the point of destination. Attractive features of these nanoparticles include extremely large surface areas associated with large pore volumes, adjustable pore sizes between 2 and about 20 nm, and great flexibility regarding the introduction of molecular functionalities in the pores and on the external particle surface. Particle sizes from about 100 nm to 300 nm are deemed optimal for cellular uptake by endocytosis. Additionally, the internal particle surface can be adapted by implementing basic, acidic or hydrophobic residues that are compatible with the cargo of choice. Surface functionalization can also be used to implement docking points for targeting ligands or dye labels enabling theranostic applications.

While biocompatibility, cell uptake and effective delivery of therapeutic model cargos such as dye molecules were initially in the focus,[1] in recent years MSNs have been investigated for different specific medical applications,[2-4] including cancer therapy.[5] In this context, the delivery of short interfering RNA (siRNA) raises hopes for achieving highly specific anti-cancer treatments with low side-effects. The siRNA molecules are double stranded short chain oligonucleotides with 20-23 base pairs that post-transcriptionally regulate protein synthesis by sequence-specific matching with mRNA molecules and thus triggering their degradation. The discovery of RNA interference in 1998 by Fire and Mellow, awarded with the Nobel Prize in 2006, has set off a surge in research activities trying to exploit synthetic siRNA as a therapeutic tool with first clinical trials underway.[6, 7] However, due to factors including their small size (leading to effective clearing from the body), their negative charge that is incompatible with the cell membrane, and their limited chemical stability, siRNA molecules have to be packaged or stabilized in order to successfully enter target cells via endocytosis. Several different packaging schemes have been investigated including the use of polymers, liposomes, dendrimers, hydrogels or inorganic host systems.[6, 8]

Mesoporous silica nanoparticles are viewed to be promising candidates for stabilizing and packaging siRNA because they offer a chemically robust matrix and a pore system that can, in principle, accommodate even small oligonucleotides. However, due to the large molecular size of siRNA of approximately 2 X 8 nm and a hydrodynamic diameter of about 4.2 nm, sufficiently large pore entrances are necessary for an efficient uptake of siRNA molecules in the MSN pore system. Typical MSNs feature pore sizes at around 3 nm, and as a result such MSNs have been used to accommodate siRNA on their external surface, usually aided by large cationic polymers such as polyethyleneimine (PEI)[9-11] or poly(2-dimethylaminoethyl methacrylate) (PDMAEMA)[12]. These polymers are known as endosomal escape agents, attributed to a proton-sponge effect, however, they are also known for their toxicity and are for this reason sometimes used in combination with polyethylene glycol (PEG) to avoid these side effects.[13, 14]

In order to offer improved protection as well as a higher loading capacity for siRNA, an uptake into the internal pore system of silica nanoparticles is highly desirable. Very few examples have been published where silica nanoparticles with pores larger than 10 nm were used as carrier system for siRNA. In an early study, Gao et al. showed that a new class of large-pore MSNs (LP-MSN) with 20 nm pores was able to absorb plasmid DNA and protect it against enzymatic degradation.[15] Ashley et al. used microemulsion-derived LP-MSNs with 13-30 nm multimodal pores in combination with a supported lipid bilayer to host a 10-100-fold amount of siRNA as compared to liposomal carriers and showed their good efficacy against translation of different cyclin proteins.[16, 17] Na et al. synthesized sponge-like structured silica particles with about 23 nm pores that were subsequentially aminated and decorated with PEG. These particles, loaded with siRNA, efficiently reduced the green fluorescent protein (GFP) and the vascular endothelial growth factor (VEGF) in Hela cell xenografts upon direct injection.[18] Other examples that used large-pore carriers for siRNA delivery needed the assistance of PEI to show good efficacy in gene knockdown.[19, 20] Lin et al. used 10 nm LP-MSN and cross-linked PDMAEMA via disulfide bonds to the host in order to achieve an intracellular cleavage of the polymer and a triggered siRNA delivery. They achieved silencing of luciferase as well as of the endogenous protein Lamin A/C of about 60% with low siRNA loadings of about 3 wt%.[21]

The group of Gu demonstrated that it is not a prerequisite to use extra-large pores in order to adsorb siRNA into the internal pores of silica nanoparticles. They used chaotropic salt solutions to shield the repulsive negative charges between silica surfaces and nucleic acids and were able to adsorb up to 27 µg/mg siRNA into the 3.7 nm pores of unfunctionalized MSN.[22] These particles showed knockdown efficiency against GFP and VEGF when PEI coupled to the fusogenic peptide KALA was attached on the outside of their particles.[23, 24]

Here we describe the development of a novel platform of medium-pore core-shell porous silica nanoparticles aimed at a high siRNA loading and release capacity as well as high siRNA knockdown efficiency. Internal siRNA uptake is achieved by using core-shell silica particles exhibiting a positively charged interior due to amino groups and a negatively charged exterior based on sulfhydryl groups to minimize external adsorption. In previous studies we could show that core-shell functionalized small-pore MSNs were efficient carriers for drug delivery of small molecules.[25] Co-condensing functional groups in the particle core and orthogonal groups on the particle surface allows for an efficient implementation of various functionalities such as particle labeling, attachment of pore-closing molecules and specific targeting agents in a spatially controlled manner. this strategy was successfully used for the selective delivery of cisplatin to cancerous cells in lung cancer tissue while sparing healthy cells.[26]

In this work we show that co-condensation derived core-shell nanoparticles with medium-sized pores (MP-MSN) of 4 nm can be charged with an extremely high load of up to 380 µg/mg siRNA/MSN even without using highly concentrated salt solutions. The subsequent siRNA desorption was quantitatively evaluated as a function of structural and process parameters in order to optimize the elution efficiency of siRNA for later applications in cell experiments. To enhance the potential for siRNA delivery, we developed a novel strategy for cell transfection by combining MSN particles with modularly designed block copolymers. The polymer presented here and more complex analogs of these modular polymers have been used successfully by some of us for siRNA delivery through polyplex formation.[27, 28] siRNA polyplexes often suffer from very variable, heterogeneous nanoparticle sizes. In the present work, polymers are attached to the exterior of our approximately 150 nm homogeneous MSN particles to inhibit premature release and to allow for endosomal escape after transfection into KB cells. These polymers contain segments with multiple succinoyl tetraethylene pentamine (stp) that can interact with the negatively charged silica surface. Lysine is also used as a branching point for attachment of oleic acid units to establish hydrophobicity and membrane permeability.[29] This combination of multifunctional MSNs and modularly designed block copolymers proved to be highly efficient in the down-regulation of a GFP-luciferase fusion reporter protein.

**Results and discussion**

**1 Particle synthesis and analysis**

Mesoporous silica nanoparticles have been proposed as host materials for truly internal adsorption of nucleic acids since 2006, when Solberg et al. used micron-sized mesoporous silica to adsorb linear DNA with 760 to 2000 bp.[30] To facilitate the adsorption of negatively charged, large plasmid DNA (pDNA) molecules of over 5000 base pairs, Gao et al.[15] adapted a new synthesis route that resulted in exceptionally large pore diameters between 12 and 20 nm while maintaining particles sizes between 70-300 nm by using a fluorinated polymer FC-4 as particle growth inhibitor. The silica surface was subsequently grafted with aminoproylsilane (APTES) in order to create a cationic layer that could interact with the phosphate moieties of the pDNA. A good protection against endonucleases was demonstrated through this encapsulation. Porous hosts that are able to internally adsorb nucleic acid molecules offer both protection from external attack and a large pore volume that promises high loading capacities. This is expected to be advantageous compared to electrostatic attachment on the outside of inorganic particles, as it is often performed for RNA delivery with small-pore MSNs.

Here we investigated functionalized mesoporous silica hosts with large pore sizes of about 10 nm (LP-MSN), in addition to nanoparticles with medium-sized pores between 4 and 5 nm (MP-MSN). The LP-MSNs were made by a multistep acid-catalyzed synthesis using F123 as a template accompanied by trimethylbenzene as swelling agent and FC-4 as particle growth retardant. These purely siliceous particles were subsequently grafted with APTES or a mixture of APTES and phenyltriethoxysilane (PhTES), resulting in a homogenous distribution of amino groups throughout the particle body. Mesoporous particles sized between 100 and 200 nm are formed as shown in the transmission electron micrograph (TEM) in Figure 1a. TEM and nitrogen sorption data in combination indicate that the pores have a bottleneck morphology. When the pore-size distribution is analyzed by using the adsorption branch we observe a size-distribution around 11 nm. However, when the NLDFT-equilibrium model is used we observe a pore-size distribution around 7.8 nm reflecting the presence of smaller pore windows (see figure 1b).

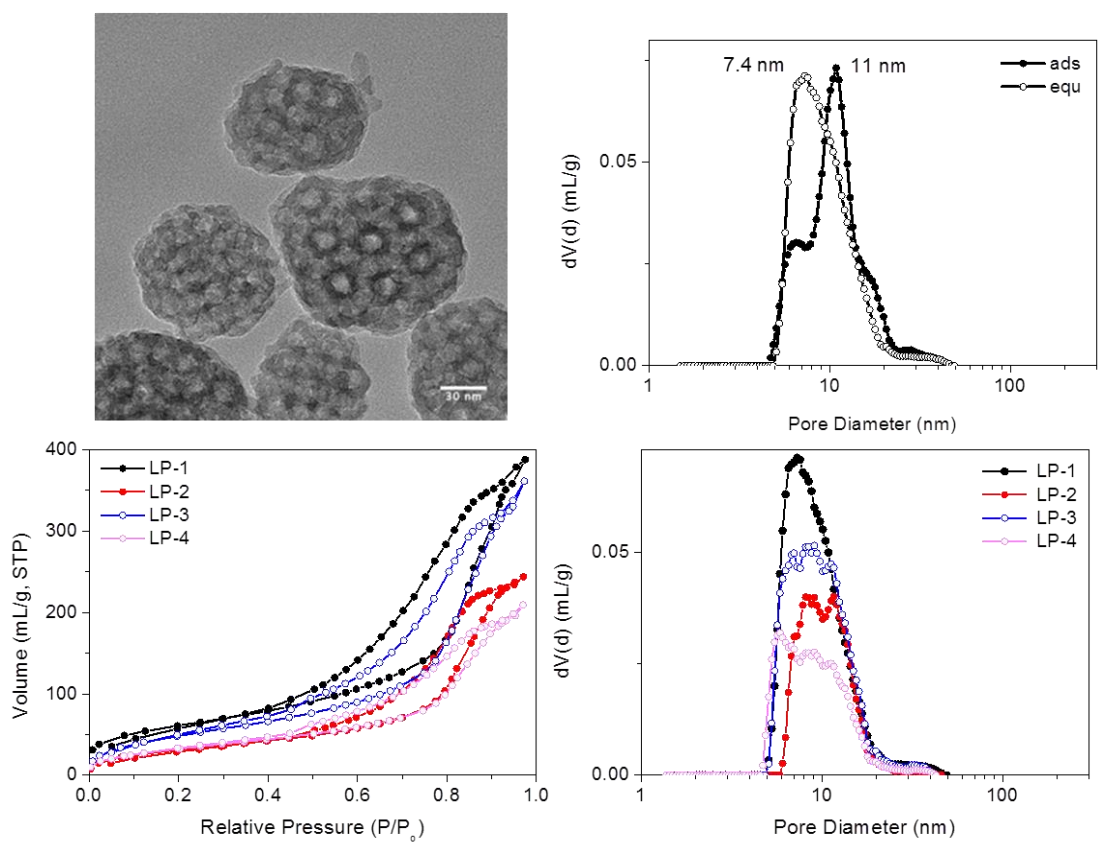

Figure 1: Characterization of large-pore silica nanoparticles (LP-MSN). a) Transmission electron micrograph  b) pore-size distribution of sample LP-1 derived from the NLDFT adsorption branch model and NLDFT equilibrium modelof the respective nitrogen sorption isotherm c) nitrogen sorption isotherms derived from parent siliceous sample LP-1 and samples LP-1 to LP-4 after grafting of different amounts of PhTES and APTES (see Table 1), and d) corresponding pore-size distributions derived from the NLDFT equilibrium kernel for cylindrical pores.

Due to their large pore size and corresponding wall thickness, the parent LP-MSNs show a relatively small surface area of 220 m$^2$/g, which is further reduced to 120 m$^2$/g after grafting with increasing amounts of phenyl and amino residues. The pore volume is concomitantly reduced as shown in the associated pore-size distributions (Figure 1d).

In addition to LP-MSN, we created MP-MSN nanoparticles with a pore size sufficient for small oligonucleotide adsorption and with orthogonal surface functionalities that drive the siRNA into their inner void volume. For this purpose, a core-shell co-condensation reaction was performed that results in a positively charged interior lining as well as a negatively charged outer shell. Medium-sized pores were obtained by using a modification of our previously established synthesis route.[31-33] In brief, a preheated solution containing cetyltrimethylammonium chloride (CTAC) as template was enriched by the pore-expanding agent triisopropylbenzene (TiPB) and mixed with a preheated solution containing tetraethoxysilane (TEOS) and APTES at elevated temperature for 20 minutes, followed by addition of the ingredients for the second particle layer (containing TEOS and mercaptopropylsilane (MPTES)). The amounts of the silane coupling agents were varied to achieve different relative surface concentrations of amino- and mercapto-groups.

Nanosized MP-MSN particles of around 150 nm were obtained as shown in the transmission electron micrographs in Figure 2 for the different particle compositions. These were either 9 mol% amino groups (of the combined silica precursors) in the core and 1 mol% mercapto groups in the shell or 4 mol% amino groups in the core and 2 mol% mercapto groups in the shell (see also Table 1). Surface areas varying between about 900 and 700 m$^2$/g reflect the different concentrations of functional groups present, with the smaller amount of aminopropyl groups in sample MP-3 resulting in a higher surface area as well as a larger pore volume. A comparable pore size of 4 nm was obtained in sample MP-3 and MP-2 despite changes in composition, while a pore-widening to 4.7 nm was achieved in sample MP-1 by changes in sample treatment before template extraction (see Experimental Part). The pore morphology can be described as an ordered stellate arrangement with conically widening mesopores, with pore-size distributions derived from the adsorption and desorption branches being nearly identical.

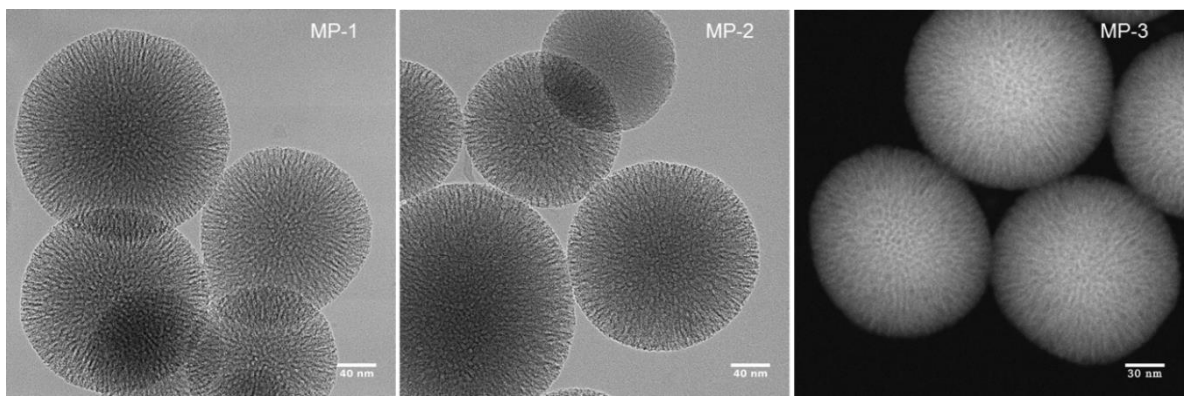
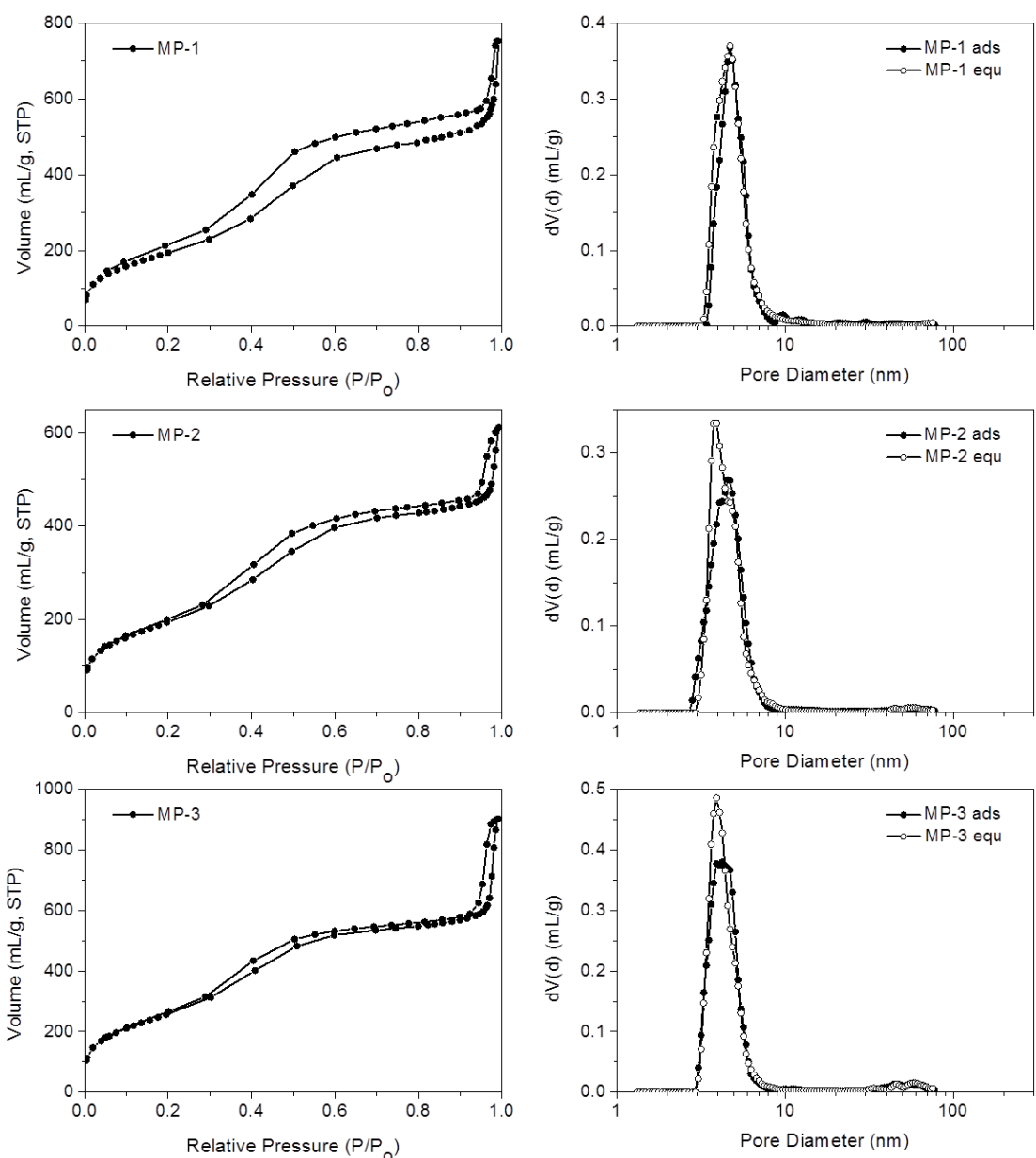

Figure 2: Transmission electron micrographs (MP-1 and MP-2) and scanning transmission electron micrograph (MP-3) of core-shell MP-MSN and the corresponding nitrogen sorption isotherms with pore-size distribution curves derived from the NLDFT equilibrium kernel for cylindrical pores (the NLDFT adsorption branch model and NLDFT equilibrium model superimposed).

Table 1: Sample compositions and surface properties

| Sample | Composition (mol%) | Surface area (m$^2$/g) | Pore size (Cavity/pore) (nm) | Pore volume At 0.8 p/p$_0$ (ml/g) | Particle size (nm) (TEM) | IEP (pH) |
|---|---|---|---|---|---|---|
| LP-1 | pure silica | 220 | 11/7.4 | 0.60 | 70-170 | 4.0 |
| LP-2 | 9% NH$_2$ | 120 | 8-12/9.9 | 0.38 | | 10.1 |
| LP-3 | 4% Ph 6% NH$_2$ | 186 | 7-12/9.8 | 0.56 | | 9.9 |
| LP-4 | 6% Ph 3% NH$_2$ | 123 | 6-13/9.6 | 0.32 | | 8.6 |
| MP-1 | 9 % NH$_2$ 1% SH | 670 | 4.7 | 0.75 | 150 | 5.9 |
| MP-2 | 9 % NH$_2$ 1% SH | 694 | 3.9 | 0.66 | 150 | 6.3 |
| MP-3 | 4% NH$_2$ 2% SH | 937 | 4.0 | 0.85 | 150 | 4.8 |

**2 SiRNA adsorption and release studies with LP-MSN**

A prerequisite for any efficient siRNA delivery system is a sufficient loading capacity and high desorption yield, as well as an adequate rate of release of the cargo under conditions that resemble cellular environments. To investigate the behavior of our different LP-MSN and MP-MSN carrier systems, we undertook siRNA adsorption and desorption studies under varying conditions.

It has been assumed that the relatively large siRNA oligonucleotides require a minimum pore size of 5 nm or more for unhindered diffusion into the particle interior.[15] Furthermore, it is recognized that the negative charge of siRNA prohibits its inclusion into siliceous materials that are negatively charged at neutral pH, and that a surface decoration with positively charged groups is necessary. In order to introduce surface amino groups into LP-MSN we performed a post-synthetic grafting with APTES since a direct cocondensation of more than 1 mol% APTES disrupts the silica network under the acidic reaction conditions. We used APTES in combination with PhTES as additional shielding against silanol surface groups. This way we were able to change the surface properties of LP-MSN systematically, as reflected in the zeta potential curves in Figure 3a. The isoelectric point (IEP) shifts over the whole range between pH 4 to pH 10. In Figure 3a, we included our purely siliceous LP-MSN parent particles for comparison (LP-1). These siliceous mesoporous particles have an isoelectric point around 4, thus at neutral pH they feature a negative charge of -10 to -20 mV that prevented any significant adsorption of siRNA.

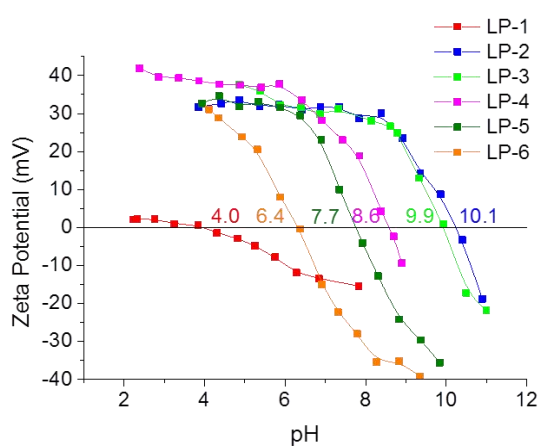
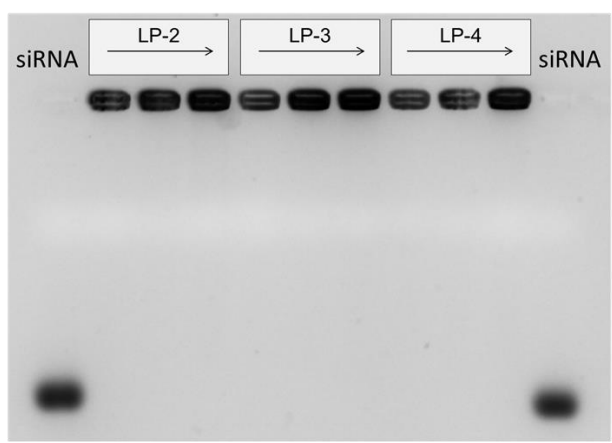

Figure 3: (a, left) Zeta potential measurements of LP-MSN with varying surface functionalization, (b, right) gel electrophoresis measurements with samples LP-2, LP-3 and LP-4 loaded with either 2.5, 6.6, or 12 µg siRNA/mg MSN (increasing concentration from left to right)

When 0.5 mol% amino groups were included into the internal particle surface in addition to 2.5 mol% phenyl groups the IEP was shifted from 4.0 to 6.4 (LP-6). However, siRNA adsorption was only minimal. For any significant amount of adsorption, a 1 /1 mol% amino/phenyl content was necessary, having an IEP of 7.7 (LP-5). Higher siRNA loadings were achieved with amino-group contents exceeding 1 mol% throughout the particle body. Increasing the APTES content from 3 to 9 mol% in LP-4 to LP-2 we obtained particles with very high IEP´s and thus high positive zeta potentials at pH = 7 between 30 to 40 mV. These surface charges were sufficient for an uptake of siRNA between 32 to 40 µg siRNA/mg MSN. Surprisingly, when the desorption process was studied by exchanging the solution for a fresh aqueous solution, nearly no elution was observed even after several hours under stirring. However, offering a cytosol-simulating PBS buffer solution at pH 7.4, the release was slightly improved to about 5% RNA after 1 h, or 23% after 1 day (fraction of loaded RNA released; see Supplemental Information, SI-S1). This tight binding of siRNA in these samples is reflected in gel shift results (Figure 3b). Here, samples LP-2, 3 and 4 were each loaded with 2.5, 6.6 or 12 µg siRNA/mg MSN. The siRNA was retained tightly in all samples even without the presence of any pore-capping systems present on the particles. No siRNA elution is visible when 100 V was applied for 1.5 h to the gel.

Similar difficulties regarding the release of pDNA from comparable particles were encountered already by Gao et al.[15], who used a 2M NaCl solution at 50°C to free the DNA, conditions that are clearly not compatible with cell experiments. Only one other research group used these bottleneck LP-MSN and modified them with either APTES or by covalently attaching polylysine to the surface. This way, Hartono et al. increased the loading capacities with RNA-mimicking oligo-DNA to 57 µg/mg. Nevertheless, when applied in cell experiments with functional siRNA, only low efficacies of 15 or 30% were reached when using very high siRNA concentrations of 100 nM.[34] In follow-up studies they adsorbed the endosomolytic agent chloroquinone into the pores while siRNA was coordinatively attached via poly(2-dimethylaminoethyl methacrylate) (PDMAEMA) on the outside of the silica particles.[19] Good efficacies were still only reached when the LP-MSNs were equipped with covalently bound PEI instead of PDMAEMA.[35] In light of these collective findings we suggest that not only the pore size and surface charge are important in determining the loading capacity, but that the pore morphology might play a pivotal role in efficient siRNA release. Nitrogen sorption data of LP-MSN show a bottleneck-type pore morphology that features smaller openings than pore diameters, which might compromise an efficient diffusion of the highly charged molecules. These examples also show the importance of studying the siRNA elution process as a basis for choosing the appropriate hosts for siRNA.

**3 SiRNA adsorption and release studies with MP-MSN**

For reasons discussed in the previous section, we developed and evaluated stellate-morphology medium-pore (MP-MSN) samples for loading and release of siRNA. We offered siRNA at different concentrations to aliquots of MP-MSN samples and followed the uptake after 15 minutes and 1 h by measuring the remaining siRNA concentrations in the supernatants. Similarly, we eluted the adsorbed RNA by removing the supernatant and exchanging the solution by equal amounts of either water or a PBS buffer solution at pH = 7.4. Figure 4 compares results obtained with samples MP-1 and MP-2 having slightly different pore sizes of 4.7 and 4 nm, respectively.

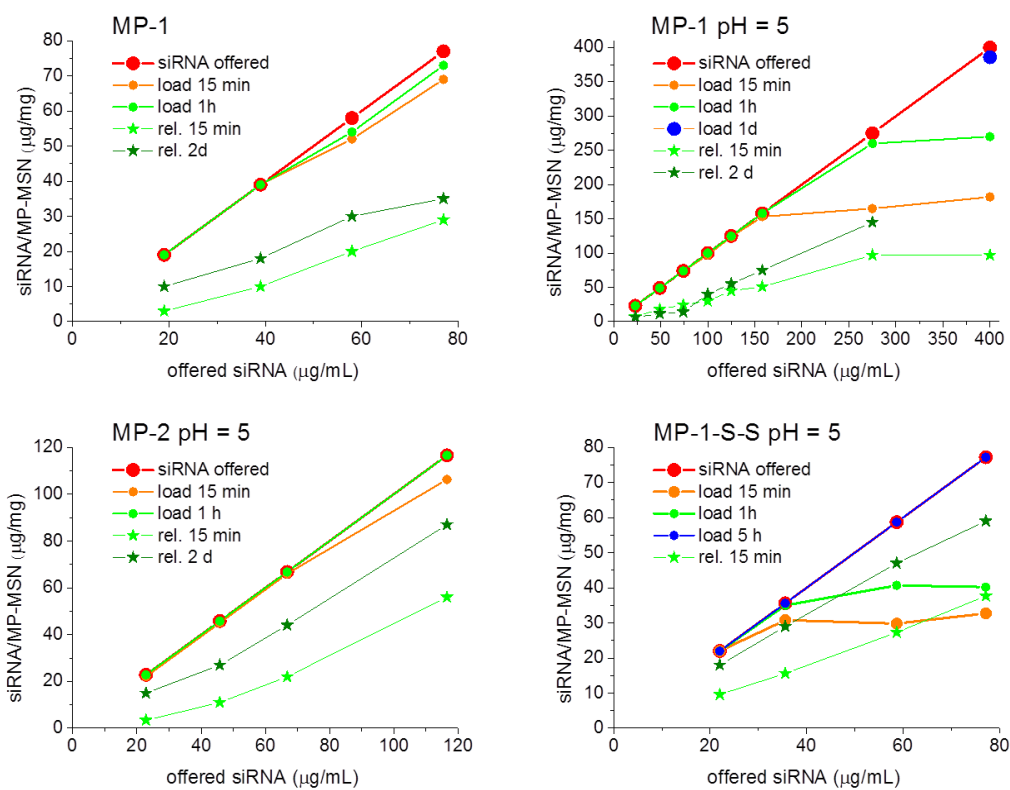

Figure 4: siRNA uptake and release studies with stellate MP-MSN. Aliquots of 100 µg of MSN were exposed to 100 µL of siRNA solutions with increasing concentrations. (a, top left) Uptake in MP-1 from water, and (b-d, top right and bottom) uptake from MES buffer at pH = 5 into samples (b) MP-1, (c) MP-2, and (d) MP-1-S-S. Release experiments were performed in 100 µL PBS buffer at pH = 7.4 in all samples. The red curve indicates the offered siRNA amount, and orange and green dots indicate the adsorbed siRNA after 15 minutes and 1 h, respectively. Light green and dark green stars are the cumulatively released amounts of RNA after 1 h and 2 days, respectively.

The siRNA amount offered is indicated by the red graph in Figure 4, and the actual siRNA uptake after 15 minutes and after 1 h was color-coded with orange and green dots, respectively. The eluted siRNA amounts are depicted in green colored stars after the indicated time. It is evident from the data in Figure 4 that in sample MP-1 with pores of only 4.7 nm diameter a rapid uptake of siRNA within 15 min takes place up to a particle loading concentration of 40 µg siRNA/mg when the siRNA is offered in an aqueous solution. Even larger amounts to at least 78 µg/mg are rapidly adsorbed after 1 hour of exposure. When siRNA is offered in an MES-buffer solution at pH = 5 (MP-1, pH = 5), the uptake after 15 min can be increased to 150 µg RNA per mg MSN. At very high siRNA concentrations, the adsorption slows down. However, even when 275 or 400 µg/mL RNA is offered as loading solution, a nearly complete uptake occurs after 1 h exposure in the former case, while a very high loading of 386 µg/mg is reached after 1 day. To our knowledge, these are the highest amounts of siRNA that have been included in such narrow-pore nanoparticles. A rapid and efficient adsorption takes also place in the sample MP-2 with narrower pores of only 4 nm, which when exposed up to 120 µg/mL siRNA solution showed a complete adsorption after 1 hour.

Since an equally important feature for later cell experiments is the efficient release of siRNA, we studied these samples after stirring in a PBS buffer solution at pH = 7.4 and followed the elution in time. As shown with the green-starred graphs in Figure 4, the sample MP-1 eluted between 16 and 40 % of the adsorbed siRNA within 15 min and up to 55 % after 2 days. In contrast, the narrow-pore sample MP-2 eluted a remarkably higher content of siRNA after 2 days amounting to between 66 and 75 %, depending on the siRNA loading.

To also achieve a higher release of siRNA in sample MP-1, we coupled amino groups in MP-1 via a redox-reactive disulfide bridge to the particle body. Such a procedure was proposed by Zhang et al., who argued that a stimuli-responsive coverage of the inner and outer surface could result in a triggered and thus more efficient release of small oligonucleotides.[36] They transformed the amino-covered surfaces of MP-MSN into disulfide-coupled amino groups that are then cleavable under the reductive reaction of glutathione (GSH), which is present in concentrations of up to 10 mM in the cytosol. They observed the highest release of oligo DNA under the action of GSH with particles having 4.5 nm pores. We adapted their procedure in order to compare the release behavior and finally the transfection efficiency between our samples with and without disulfide-coupled amino groups. Thus, in a first step surface amino groups were converted via succinic anhydride coupling into carboxy groups. These were subsequently reacted with cystamine catalyzed by the EDC /NHS-sulfo reaction to result in amino-terminated cleavable residues (MP-1-S-S). The progress of this reaction was confirmed by FTIR spectroscopy and zeta potential measurements, as shown in the supplemental material (see SI-S2).

RNA adsorption into MP-1-S-S turned out to be markedly slower than in the purely co-condensed sample MP-1, most likely due to the pore size reduction caused by the long residual linker. In this case, instead of 1 hour it took 5 hours for a complete uptake of the administered siRNA when measured with concentrations of up to 80 µg/mL (see Figure 4, MP-1-S-S). On the other hand, the desorption was more efficient, showing a release of 43-49% already after 15 minutes and between 76 and 81% after 1 day (depending on siRNA loading). Notably, no reducing agent was used here for release.

We believe that the observed differences in diffusivity are induced to a large part by electrostatic interactions. On the one hand, the higher the concentration of protonated amino groups on the internal surface the more efficient and faster is the adsorption behavior. Purely siliceous mesoporous silica is not able to adsorb siRNA, and neither did we observe any significant uptake when the surface in LP-MSNs was covered with cyano, glycidoxy or with mercapto groups. More than 1 mol% amino groups are necessary for any significant uptake of siRNA. Additionally, pore size and pore morphology influence the distribution of short oligonucleotides with respect to the pore walls. In LP-MSN a side-by side adsorption of RNA molecules may be feasible by the large size of the entrance pores. The pore-size in MP-MSNs is close to the hydrodynamic radius of siRNA and a single-file diffusion is more likely to occur. Reducing the pore-size in these samples brings the RNA molecules into a closer contact with the pore surface and its positive charge so that the progressing adsorption is slowed down. This might explain the differences between MP-1 and MP-2, which have a difference in pore-size of 0.7 nm. Similarly, by introducing additional functional groups to the particle surface as in MP-1-S-S, the pore size is effectively reduced and the uptake of siRNA is again slowed down.

However, even a wide pore size allowing for facile accommodation of the large guest molecules does not guarantee efficient release kinetics. In very large pores the association of the siRNA with the charged surface is likely only partial or one-sided and overall adsorption is possibly less ordered. Entanglement and clustering of the stiff siRNA molecules is likely when buffer solutions are added. This may be a reason for slower desorption, which appears to be more pronounced with a bottleneck pore morphology, as encountered in sample LP-1-3. In contrast, a tighter fit of the surrounding pore surface with the highly negatively charged phosphate groups of the double stranded RNA and single file diffusion in the straight channels of MP-MSN apparently is advantageous for efficient release once a sufficient ionic strength is present in the reaction medium. This ease of desorption is further increased when the internal charge of the MSN is lower, such as in sample MP-3. With only 4 wt% amino groups in the sample core we encounter a fast and efficient desorption of 60 to 70% of the adsorbed siRNA already after 15 minutes (see SI, S3a).

**4 Cell transfection and GFP-luciferase knockdown**

Since a very high siRNA uptake and efficient release behavior was established in this work, we chose samples MP-1-S-S and MP-1 as candidates for comparative siRNA delivery studies with KB-cells. To reduce untimely elution of siRNA and simultaneously add endosomal escape functionality to our system, we combined the siRNA-loaded MP-MSN particles with the modularly designed block copolymer 454.[28] It is composed of multiple succinoyl tetraethylene pentamine (stp) units that carry three basic amino groups, while the terminal amino and carboxy groups are used for coupling to additional lysine (K) or tyrosine (Y) units. The latter are further terminated with cysteine modules that potentially can undergo disulfide bridging through their mercapto groups or could be used for adding additional functionality such as targeting modules. Lysine groups are used as branching points for the attachment of oleic acid segments, which are included for establishing endosomolytic properties (see Scheme 1). This feature is especially important since most of the delivering vectors are taken up via endocytosis. However, if a timely escape from the endosome cannot be established degradation sets in and the delivery will fail.

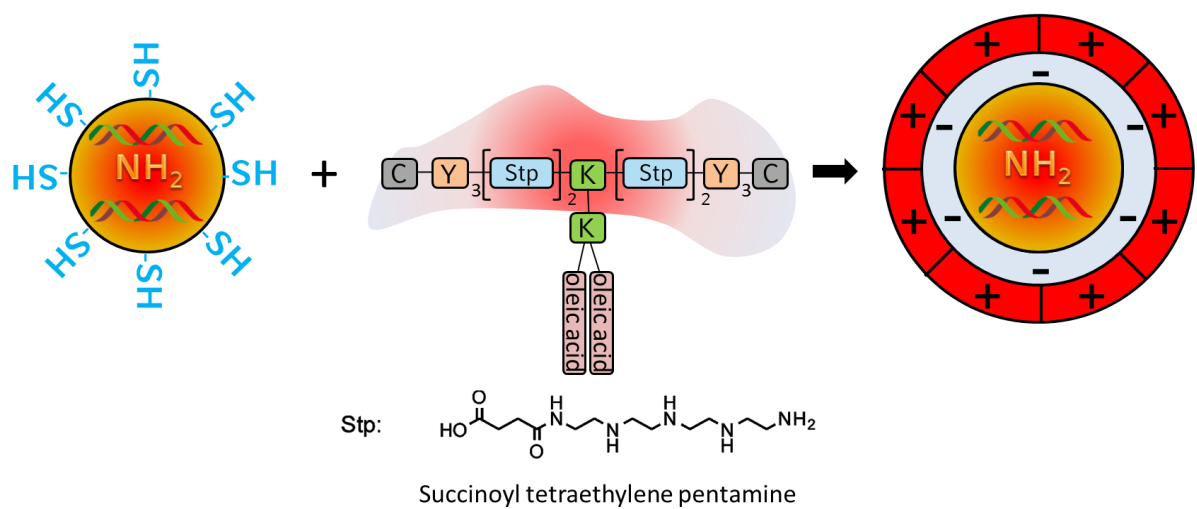

Scheme 1: Construction of an siRNA delivery platform by combining multifunctional core-shell nanoparticles with modularly designed polymer 454.

Attachment of this cationic polymer to the siRNA-loaded MSN particles occurs likely through electrostatic interactions with sample MP-1 as well as in MP-1-S-S, even though the surface layer is different in both cases. In MP-1-S-S, the originally mercapto-terminated outer surface is completely transformed into amino groups through reaction with the cystamine molecules. This was established by Raman spectroscopy and zeta potential measurements and is detailed in the SI in Figure S4. Thus, in this sample some of the adsorbed siRNA was additionally bound to the external surface, rendering it negatively charged again. Coulomb interactions with the cationic polymer are thus possible in both samples, and are confirmed by zeta potential measurements (see SI; Figure S5). Initially, the IEP in sample MP-1 was relatively high (5.9) since the high loading with amino groups was not completely screened by the 1 mol% terminating mercapto groups in the particle shell. This effect was partially neutralized after siRNA adsorption and the IEP shifted to a lower value of 5.1. However, upon exposure to the cationic polymer the surface charge increased again and the IEP reached a value of 6.4. This is still much lower than the IEP of 7.6 of the pure polymer, confirming its binding to the particle surface.

Cell transfection with these siRNA delivery systems was next verified with confocal microscopy, shown in Figure 5. Wild-type KB cells were first incubated for 45 minutes with sample MP-1 and unbound particles were removed by a medium exchange. Microscopy was then performed after 24 h and 48 h. To visualize the internalization of siRNA, we labeled 20% of the siRNA with the dye CY5 before adsorbing it into the carrier MP-MSN. A significant uptake of siRNA is observed in the green stained KB cells at both times (see top row in Figure 5). To track the location of the siRNA within the MSN particles, we labeled the MP-MSN carrier to some degree with NHS-ATTO-488 before adsorbing the partially labeled siRNA. A Z-stack maximum projection is shown in the middle row in Figure 5 illustrating the large uptake of MSN particles in the cells. The green cell membrane staining was omitted here to optimize the visibility of the green MSN particles and the red siRNA. Significant co-localisation between MSN carrier and siRNA is visible both after 24 h and 48 h. We note that the released fraction of siRNA is not observed due to dilution effects. In the bottom row of Figure 5 we show micrographs in which the particles were labeled green again with NHS-ATTO-488, but now part of the polymer was labeled with Mal-ATTO-633 to confirm the sustained attachment of the polymer to the particle surface. This is confirmed through the yellow (additive) coloring.

Further cell experiments were undertaken to study the biological activity of the MP-MSN carrier system. If not stated differently, we used the same siRNA loading of 5 wt% siRNA for this purpose as for the cell microscopy studies. First, the complete uptake of siRNA into the MP-MSN host was confirmed by measuring the supernatant solution, then samples were redispersed and the block copolymer was added at increasing amounts to evaluate the best transfection conditions. After incubating the polymer for one hour at 37°C under shaking, samples were centrifuged in order to replace the supernatant with a PBS buffer solution and to remove any surplus block copolymer or free siRNA. Cells were usually exposed to 10 µg MSN/well; each well had a volume of 100 µL. Each well thus contained 31 pM siRNA or 0.5 µg. This siRNA concentration was kept constant in all experiments. Gel electrophoresis was performed with the same samples that were used for cell transfection.

In Figure 6 we show the results for sample MP-1-S-S, which was prepared using increasing concentrations of polymer/MP-MSN samples with a w/w ratio ranging from 1/100 to 1.5/1.

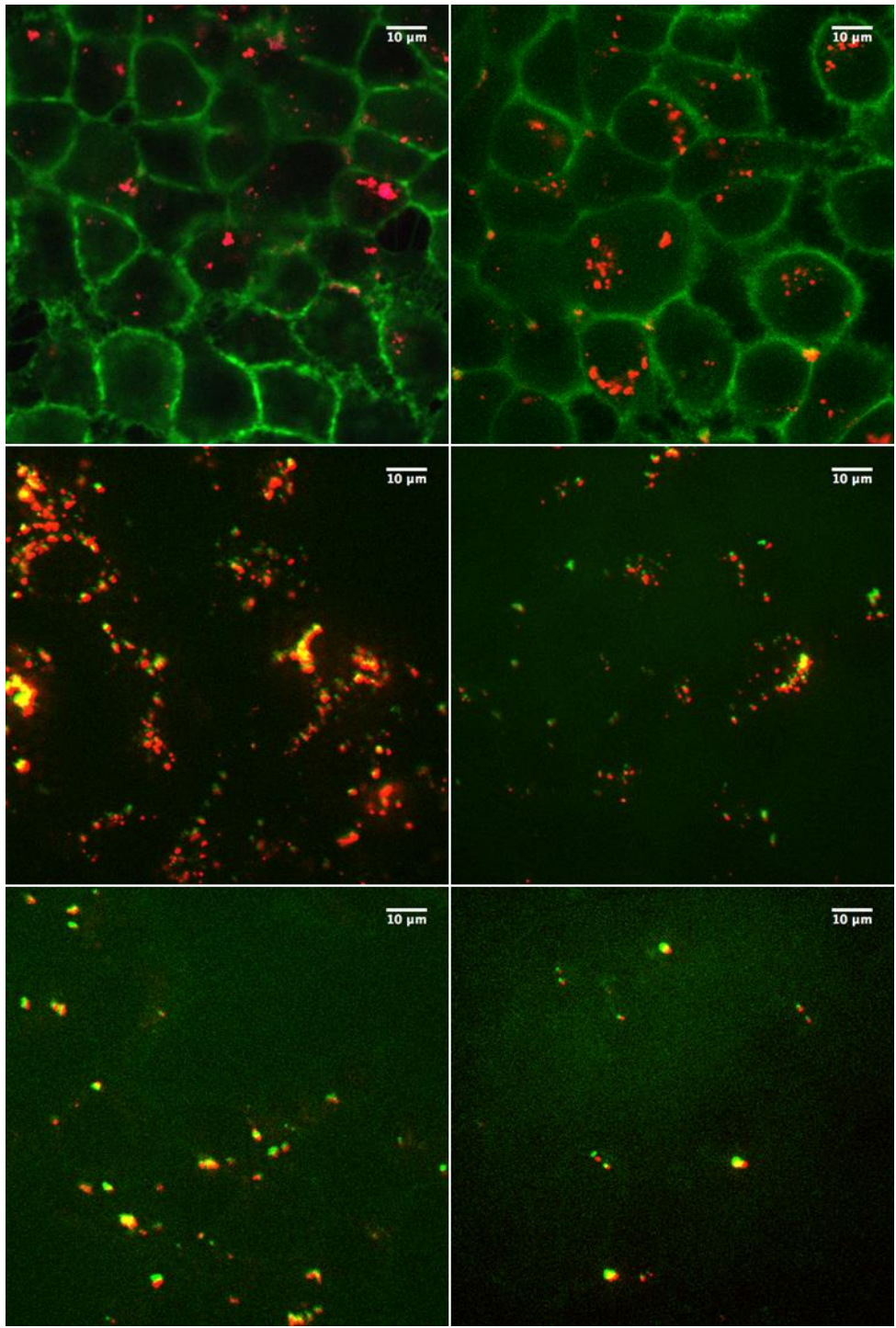

Figure 5: Confocal microscopy of the transfection of sample MP-1 in KB cells.
Left column: after 24 h incubation, right column: after 48 h incubation.
Top row: particles with partially CY5-labeled siRNA and green stained KB cells.
Middle row: Z-stack maximum projection; NHS-ATTO-488 stained MP-MSN (green) loaded with partially CY5-labeled siRNA (red).
Bottom: NHS-ATTO-488 stained MP-MSN (green) loaded with unlabeled siRNA and partially Mal-ATTO-633 labeled polymer (red).

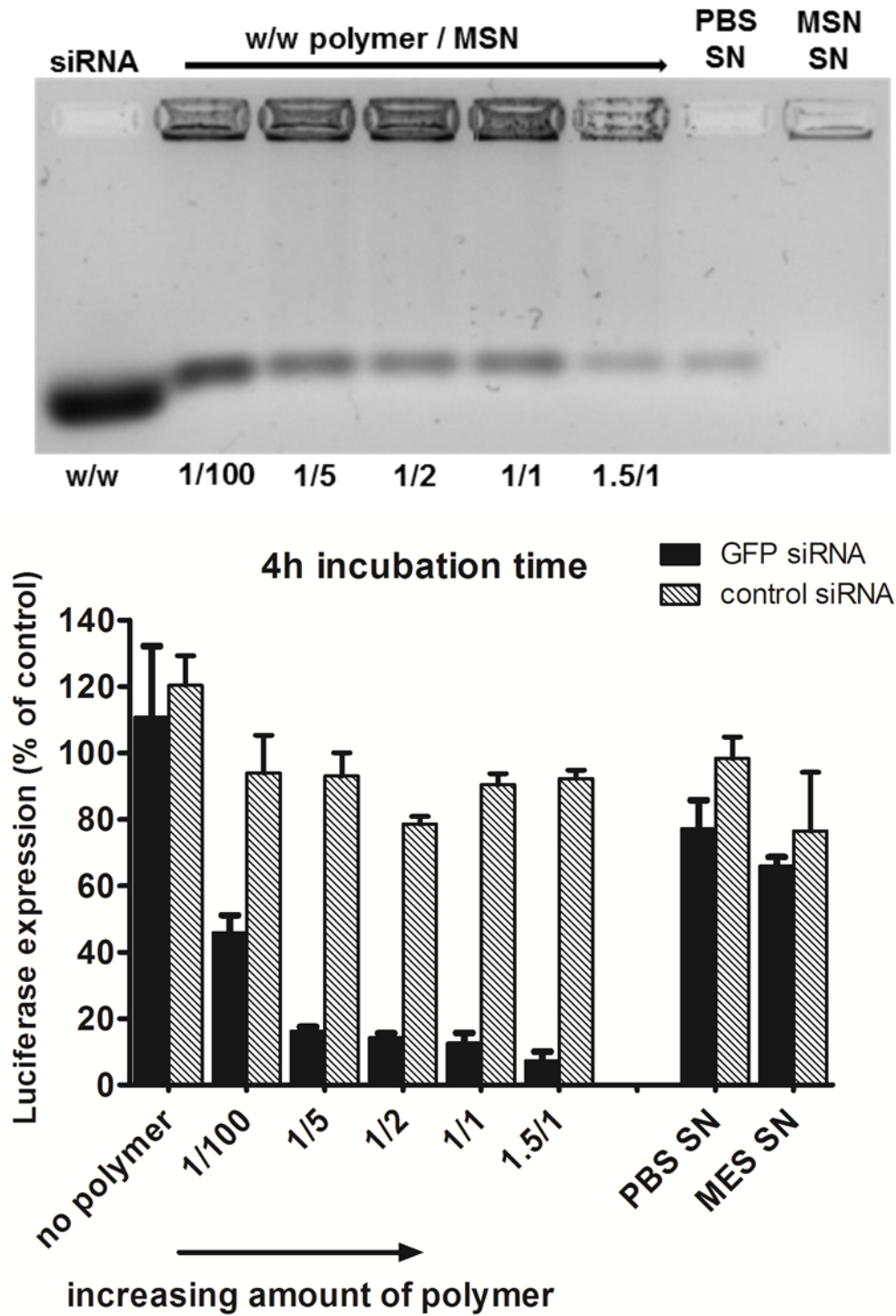

Figure 6: (a, top) Gel electrophoresis of sample MP-1-S-S with an increasing amount of block copolymer ranging from 1/100 up to 1.5/1 w/w polymer to particle. The first and the second supernatant solutions (SN) after sample preparation are included (5 wt% siRNA).
(b, bottom) The corresponding KB/eGFPLuc cell transfections after a particle incubation time of 4 h on cells, luciferase read out after 48 h (5 wt% siRNA). A sample without polymer capping is also included for comparison.

In the gel-electrophoresis experiments (Figure 6a) we compared the release of siRNA from the MP-MSN particles as a function of polymer concentration. It can be seen that all samples do retain large quantities of siRNA, but that some minor elution of siRNA is observed, decreasing with increasing polymer content. Electrophoresis showed that the supernatants of the MES loading solution (MSN-SN) and the first wash in PBS buffer (PBS-SN) contain only negligible amounts of siRNA.

The above samples were used in experiments with KB/eGFPLuc cells which stably express a GFP-luciferase fusion reporter gene. Cells were incubated for 4 hours with samples containing either GFP siRNA (black bars, for GFP-luciferase silencing) or irrelevant control siRNA (hatched bars, no gene silencing) before they were replaced with fresh medium. The luciferase expression was measured after an additional 48h and its activity is depicted in Figur 6b. A striking concentration-dependent down-regulation can be observed, starting with an efficacy of almost 60% with the smallest polymer addition, up to over 90% knock-down with the highest polymer addition. Importantly, we note that without the polymer coating, no significant knockdown activity was observed. Between a polymer/MSN weight ratio of 1/5 and 1.5/1 the differences are only small, and for further experiments we used a 1/2 w/w ratio. We note that the supernatant solutions were also analyzed (with the nanodrop protein 280 routine), where significant quantities of the polymer were found that did not bind to the MSN particles. We included the supernatant solution in the cell experiments and observed that only a minor knockdown of 35% occurs (MES SN), and even less with the second PBS washing solution (PBS SN) containing small amounts of free siRNA as seen in the gel electrophoresis. These results clearly show that the high knockdown efficacy is brought about by a concerted action of the endocytosed MP-MSN particles releasing their siRNA cargo and the endosomolytic activity of the attached block copolymer.

These results were obtained with an siRNA-loading of only 5 wt%, and the question arises if we can take advantage of the high loading capacity of our MSNs. In Figure 7a we show cell transfection experiments where we used sample MP-1 and increased the loading concentration of RNA from 5 wt% up to 20 wt%. This way we could decrease the amount of MSN particles administered to the cells while keeping the total amount of siRNA at 0.5 µg/well. Here we used a short particle incubation time on cells of only 45 minutes. Luciferase activity was measured after 48 h as before. It can be seen that in all cases we still observe a very high efficiency that allows us to reduce the particle concentration from 100 µg/mL down to only 25 µg/mL, which translates to 2.5 µg MSN/well. To our knowledge, these efficiencies are the highest that have been achieved with mesoporous silica carriers, since usually greater particle concentrations are needed to show a substantial knockdown.[18, 20, 23] In order to investigate the limiting concentration of siRNA for good efficiency we further decreased the total amount of RNA per well (see Figure 7b). Starting from 0.5 µg RNA (32 pM) per well applied with the highest loaded sample (20 wt%), we reduced the particle/RNA amount stepwise down to 4 pM per well. It is evident that 32 pM siRNA is the optimal concentration needed for a 80-90% knock-down efficiency.

How versatile and robust is the above design concept? In the following we briefly discuss several examples where the general construction principle was varied. This relates (i) to the siRNA interactions with the interior of the MSNs, (ii) to the MSN pore size, and (iii) to the nature of the polymer coat on the MSN.

Firstly, considering siRNA-MSN interactions, we ask about the importance of internal MSN charge density. Here, sample MP-3-S-S having a lower concentration of amino groups (4 mol%) was used in combination with the block copolymer. Despite the high siRNA loading capacity (see SI, S3a), the weaker electrostatic interactions with the host lead to a more facile extraction of the siRNA from the pore system as soon as the polymer is added. Hence, these samples cannot be effective transfection agents (see SI, S3b). How important is the disulfide-bridged attachment of the amino groups for the siRNA functionality? In Figure 7 we showed the luciferase knockdown when siRNA is administered via sample MP-1, which does not contain disulfide coupled amino groups. When the particles are covered with the polymer, they again induce a successful knockdown of over 80%, showing that with the efficient MP-MSN hosts the additional reductive cleavage in the cytosol is not required for effective knockdown.

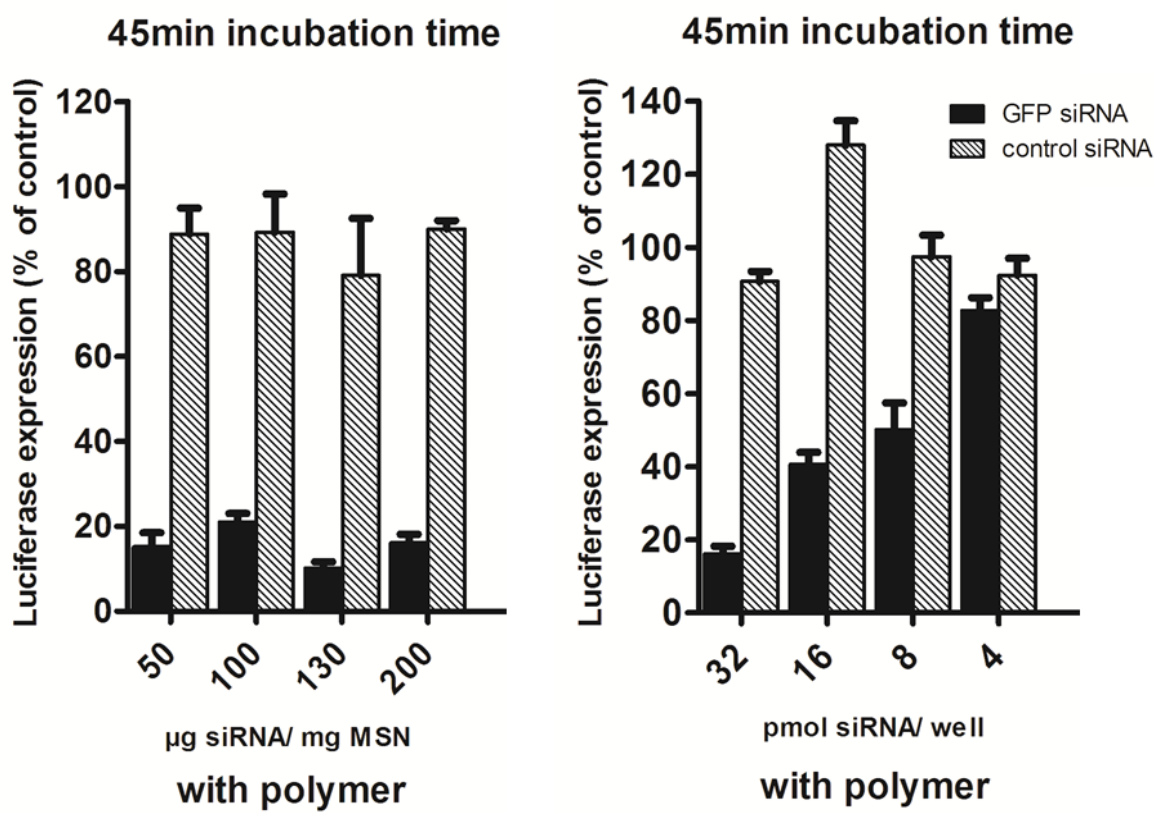

Figure 7: (a, left) KB/eGFPLuc cell transfections with sample MP-1 with increasing particle loadings of 5 wt%, 10 wt%, 13 wt% and 20 wt% siRNA, while keeping the total amount of siRNA at 0.5 µg/well (32 pM). (b, right) Cell transfections with sample MP-1 with decreasing siRNA concentrations/well. Cell transfections after a short particle incubation time of only 45 min on cells, read out after 48 h.

Secondly, we examined the impact of MSN pore size. Even with the smaller 3.9 nm pore size in sample MP-2, the same highly efficient luciferase knockdown was observed as with the larger pore system having 4.7 nm pore size (see SI, S6).

Third, we ask about the importance of the polymer capping system. We also examined the siRNA-delivery system with a standard cationic lipid by using DOTAP instead of the block copolymer. DOTAP adds a similar endosomolytic effect to our MP-MSN carrier system and shows again very high efficacies of over 80% (see SI, S7). While similar high efficacies could be achieved with this cationic lipid, it does not offer the chemical versatility and robustness of our novel block copolymer capping agent. Through its functional groups, the latter offers numerous opportunities for future covalent modification with complementary functionalities aimed at effective drug delivery.

**Conclusion**

We have introduced a new carrier system for siRNA by capping medium-size pore MSN particles with a modularly designed cationic block copolymer, resulting in excellent gene knockdown. For this purpose novel stellate core-shell mesoporous silica nanoparticles were introduced, which prove to be ideal hosts for siRNA due to their versatile surface properties. The possibility to selectively modify their internal and external surface chemistry allowed us to optimize the prerequisites for siRNA delivery. By tuning the internal surface charge as well as the pore size and morphology in the MSNs we demonstrate that the adsorption and desorption of siRNA is driven by electrostatic interactions and is best with MP-MSN featuring a stellate pore morphology. A high internal surface concentration of amino groups proved to be critical for achieving extremely large siRNA loading capacities of up to 380 µg/mg. A complementary external coverage with mercapto groups both prevents external siRNA adsorption and allows for favorable interactions with cationic polymers. In several examples we show that siRNA knockdown efficacies better than 80% can be achieved with a very low exposure of the cells to mesoporous silica (2.5 µg MSN/ 100 µL well for only 45 minutes) when the MP-MSN carriers are combined with cationic polymers. Modularly designed block copolymers serving as capping agents and ensuring endosomal escape will additionally be adaptable to specific cell environments and targeting requirements. It is anticipated that MSN particles featuring optimized morphology and internal charge density in combination with such modular block copolymer capping agents offer numerous opportunities for future *in vivo* biomedical applications of RNA-based oligonucleotides.

**Key words**: Mesoporous silica nanoparticles (MSN), core-shell functionalization, siRNA, luciferase knockdown, block copolymer

**Acknowledgements**

We thank the Deutsche Forschungsgemeinschaft (DFG) for financial support (SFB 1032). Additional support is gratefully acknowledged from the Excellence Cluster Nanosystem Initiative Munich (NIM) and from the Center for NanoScience Munich (CeNS). We owe our thanks to Philipp Klein for supplying us with the block copolymer 454.

Supporting information is available describing in detail sample preparation and analysis as well as additional experiments

**References**

1. I. I. Slowing, J. L. Vivero-Escoto, C.-W. Wu and V. S. Y. Lin, *Adv Drug Deliv Rev*, 2008, **60**, 1278-1288.
2. C. Argyo, V. Weiss, C. Bräuchle and T. Bein, *Chemistry of Materials*, 2014, **26**, 435-451.
3. Z. Li, J. C. Barnes, A. Bosoy, J. F. Stoddart and J. I. Zink, *Chem. Soc. Rev.*, 2012, **41**, 2590-2605.
4. V. Mamaeva, C. Sahlgren and M. Linden, *Adv. Drug Delivery Rev.*, 2013, **65**, 689-702.
5. N. Z. Knezevic and J.-O. Durand, *Nanoscale*, 2015, **7**, 2199-2209.
6. J. Wang, Z. Lu, M. G. Wientjes and J. L. S. Au, *The AAPS Journal*, 2010, **12**, 492-503.
7. R. Kanasty, J. R. Dorkin, A. Vegas and D. Anderson, *Nat Mater*, 2013, **12**, 967-977.
8. M. S. Draz, B. A. Fang, P. Zhang, Z. Hu, S. Gu, K. C. Weng, J. W. Gray and F. F. Chen, *Theranostics*, 2014, **4**, 872-891, 821.
9. J. Shen, H.-C. Kim, D. Kirui, J. Mai, C. Mu, H. Su, L.-N. Ji, Z.-W. Mao, F. Wang, J. Wolfram and H. Shen, *Theranostics*, 2014, **4**, 487-497.
10. T. Xia, M. Kovochich, M. Liong, H. Meng, S. Kabehie, S. George, J. I. Zink and A. E. Nel, *ACS Nano*, 2009, **3**, 3273-3286.
11. M. Wang, X. Li, Y. Ma and H. Gu, *Int J Pharm*, 2013.
12. S. R. Bhattarai, E. Muthuswamy, A. Wani, M. Brichacek, A. L. Castaneda, S. L. Brock and D. Oupicky, *Pharm. Res.*, 2010, **27**, 2556-2568.
13. W. Ngamcherdtrakul, J. Morry, S. Gu, D. J. Castro, S. M. Goodyear, T. Sangvanich, M. M. Reda, R. Lee, S. A. Mihelic, B. L. Beckman, Z. Hu, J. W. Gray and W. Yantasee, *Adv. Funct. Mater.*, 2015, Ahead of Print.
14. H. Meng, W. X. Mai, H. Zhang, M. Xue, T. Xia, S. Lin, X. Wang, Y. Zhao, Z. Ji, J. I. Zink and A. E. Nel, *ACS Nano*, 2013, **7**, 994-1005.
15. F. Gao, P. Botella, A. Corma, J. Blesa and L. Dong, *The Journal of Physical Chemistry B*, 2009, **113**, 1796-1804.
16. C. E. Ashley, E. C. Carnes, K. E. Epler, D. P. Padilla, G. K. Phillips, R. E. Castillo, D. C. Wilkinson, B. S. Wilkinson, C. A. Burgard, R. M. Kalinich, J. L. Townson, B. Chackerian, C. L. Willman, D. S. Peabody, W. Wharton and C. J. Brinker, *ACS Nano*, 2012, **6**, 2174-2188.
17. C. E. Ashley, E. C. Carnes, G. K. Phillips, D. Padilla, P. N. Durfee, P. A. Brown, T. N. Hanna, J. Liu, B. Phillips, M. B. Carter, N. J. Carroll, X. Jiang, D. R. Dunphy, C. L. Willman, D. N. Petsev, D. G. Evans, A. N. Parikh, B. Chackerian, W. Wharton, D. S. Peabody and C. J. Brinker, *Nat. Mater.*, 2011, **10**, 389-397.
18. H.-K. Na, M.-H. Kim, K. Park, S.-R. Ryoo, K. E. Lee, H. Jeon, R. Ryoo, C. Hyeon and D.-H. Min, *Small*, 2012, **8**, 1752-1761.
19. S. B. Hartono, N. T. Phuoc, M. Yu, Z. Jia, M. J. Monteiro, S. Qiao and C. Yu, *J. Mater. Chem. B*, 2014, **2**, 718-726.
20. X. Du, L. Xiong, S. Dai, F. Kleitz and S. Z. Qiao, *Adv. Funct. Mater.*, 2014, **24**, 7627-7637.
21. D. Lin, Q. Cheng, Q. Jiang, Y. Huang, Z. Yang, S. Han, Y. Zhao, S. Guo, Z. Liang and A. Dong, *Nanoscale*, 2013, **5**, 4291-4301.
22. X. Li, Q. R. Xie, J. Zhang, W. Xia and H. Gu, *Biomaterials*, 2011, **32**, 9546-9556.
23. X. Li, Y. Chen, M. Wang, Y. Ma, W. Xia and H. Gu, *Biomaterials*, 2013, **34**, 1391-1401.
24. Y. Chen, D. S.-Z. Zhang, H. Gu, X. Wang, T. Liu, Y. Wang and W. Di, *Int J Nanomedicine*, 2015, **10**, 2579-2594.
25. S. A. Mackowiak, A. Schmidt, V. Weiss, C. Argyo, C. von Schirnding, T. Bein and C. Brauchle, *Nano Letters*, 2013, **13**, 2576-2583.
26. S. H. van Rijt, D. A. Boeluekbas, C. Argyo, S. Datz, M. Lindner, O. Eickelberg, M. Koenigshoff, T. Bein and S. Meiners, *ACS Nano*, 2015, **9**, 2377-2389.

27. C. Dohmen, D. Edinger, T. Fröhlich, L. Schreiner, U. Lächelt, C. Troiber, J. Rädler, P. Hadwiger, H.-P. Vornlocher and E. Wagner, *ACS Nano*, 2012, **6**, 5198-5208.
28. C. Troiber, D. Edinger, P. Kos, L. Schreiner, R. Kläger, A. Herrmann and E. Wagner, *Biomaterials*, 2013, **34**, 1624-1633.
29. T. Frohlich, D. Edinger, R. Klager, C. Troiber, E. Salcher, N. Badgujar, I. Martin, D. Schaffert, A. Cengizeroglu, P. Hadwiger, H. P. Vornlocher and E. Wagner, *J Control Release*, 2012, **160**, 532-541.
30. S. M. Solberg and C. C. Landry, *The Journal of Physical Chemistry B*, 2006, **110**, 15261-15268.
31. K. Möller, J. Kobler and T. Bein, *Advanced Functional Materials*, 2007, **17**, 605-612.
32. J. Kecht, A. Schlossbauer and T. Bein, *Chemistry of Materials*, 2008, **20**, 7207-7214.
33. J. Kobler, K. Möller and T. Bein, *ACS Nano*, 2008, **2**, 791-799.
34. S. B. Hartono, W. Gu, F. Kleitz, J. Liu, L. He, A. P. J. Middelberg, C. Yu, G. Q. Lu and S. Z. Qiao, *ACS Nano*, 2012, **6**, 2104-2117.
35. S. B. Hartono, M. Yu, W. Gu, J. Yang, E. Strounina, X. Wang, S. Qiao and C. Yu, *Nanotechnology*, 2014, **25**, 055701.
36. J. Zhang, M. Niemelae, J. Westermarck and J. M. Rosenholm, *Dalton Trans.*, 2014, **43**, 4115-4126.

# Highly Efficient siRNA Delivery from Core-Shell Mesoporous Silica Nanoparticles with Multifunctional Polymer Caps

Karin Möller[1], Katharina Müller[2], Hanna Engelke[1], Christoph Bräuchle[1], Ernst Wagner*,[2] and Thomas Bein*,[1]

[1]Department of Chemistry and Center for NanoScience, University of Munich (LMU), Butenandtstrasse 5–13, 81377 Munich, Germany
Fax: (+49) 89-2180-77622

[2]Pharmaceutical Biotechnology and Center for NanoScience, University of Munich (LMU), Butenandtstrasse 5-13, 81377 Munich, Germany

Emails: ernst.wagner@cup.uni-muenchen.de, bein@lmu.de

**Supporting Information**

**Materials**

Fluorocarbon surfactant (FC-4, yick-vic chemicals), block copolymer surfactant (Pluronic F127, Sigma-Aldrich), 1,2,4-trimethylbenzene (TMB, Sigma-Aldrich, 98%) tetraethyl orthosilicate (TEOS, Fluka, > 98%), triethanolamine (TEA, Aldrich, 98%), cetyltrimethylammonium chloride (CTAC, Fluka, 25% in $H_2O$), mercaptopropyl triethoxysilane (MPTES, Sigma-Aldrich, > 95%), aminopropyl triethoxysilane (APTES, Sigma-Aldrich, 99%), phenyltriethoxysilane (PhTES, Sigma-Aldrich, 98%), toluene (Sigma-Aldrich, anhydrous), ammonium fluoride ($NH_4F$, Sigma, >98%), triethanolamine (TEA, Aldrich, 98%), 1,3,5-triisopropylbenzene (TiPB, Fluka, 96%), sulfo-N-hydroxysuccinimide (sulfo-NHS, Aldrich, 98%), N-(3-dimethylaminopropyl)-N'-ethylcarbodiimide (EDC, Sigma, 97%), N,N-dimethylformamide (DMF, Sigma-Aldrich, anhydrous), succinic anhydride (Fluka), cystamine dihydrochloride (Aldrich), NHS-ATTO-633 (ATTO-Tec), Mal-ATTO-633 (ATTO-Tec)
siRNA, Axolabs: GFPsiRNA: sense 5`-AuAucAuGGccGAcAAGcAdTsdT-3`, antisense 5`-UGCUUGUCGGCcAUGAuAUdTsdT-3´; ctrl siRNA: sense 5´-AuGuAuuGGccuGuAuuAGdTsdT-3´, antisense 5´-CuAAuAcAGGCcAAuAcAUdTsdT-3´
RPMI-1640, folate free (Life technologies), collagen (Biochrom), fetal calf serum (FCS, Life technologies), antibiotics (Biochrom), luciferin (Promega), lysis buffer (Promega),
oligomer 454 was synthesized as described before[28], oligomer 356 was synthesized as described before[27] GelRed™ (Biotum, Hayward, U.S.A.),
1,2-dioleoyl-3-trimethylammonium-propane (chloride salt) ((DOTAP),18:1 TAP, Avanti Polar Lipids). Millipore water was used in all experiments.

**Methods**

**LP-MSN-Synthesis and post-synthetic grafting procedures**
**LP-MSN synthesis**
The synthesis of the siliceous LP-MSN nanoparticles was performed similarly to a previous report.[15] 1.4 g of the fluorocarbon surfactant was dissolved together with 0.5 g of the block copolymer surfactant F127 in 60 g of a 0.02 M HCl solution at 60 °C under stirring at 800 rpm for 2 hours. The solution was cooled down to 15 °C in a thermostat and 0.5 g of TMB was added as pore extension agent. This mixture was stirred for 2 hours before 3 g of TEOS were subsequently added under stirring, which was continued for 24 hours. Finally, the mixture was filled into a 100 mL Teflon-lined autoclave and heated under autogenous pressure at 150 °C for another 24 hours. The product was centrifuged after cooling at 20 000 rpm for 15 minutes and the supernatant was exchanged for 30 g of a 2 M HCl solution. A second heat treatment was performed for 2 days at 140°C.

**Template extraction**
The sample was retrieved by centrifugation and the supernatant was exchanged for 35 mL of a 2 M HCl/ethanolic extraction solution and was refluxed at 90 °C for 3 hours. This process was repeated once before the sample was washed 2 times in 50 mL ethanol/water solution and was finally stored in ethanol for further use.

**LP-MSN grafting**
Grafting with APTES and PhTES was performed with varying molar ratios. Usually, 200 mg (3.3 mmol) $SiO_2$ of the parent LP-MSN sample was suspended in 20 mL of toluene under dry nitrogen in a Schlenk-flask and the silane coupling agents were added under nitrogen flow. 20 mol% (0.66 mmol) of combined silane coupling agents were added to this solution and refluxed at 90 °C for 4 hours and washed 3 times in ethanol. The following molar ratios were used for the LP-MSN samples:

LP-1: no grafting
LP-2: 20 mol% APTES (0.66 mol or 0.155 mL)
LP-3: APTES 0.44 mmol (103 µL) together with 0.22 mmol PhTES (53 µL)
LP-4: APTES 0.22 mmol (50 µL) together with 0.44 mmol PhTES (107 µL)

The resulting degree of functionalization was estimated from thermogravimetric measurements and is listed in Table 1.

**MP-MSN-Synthesis with co-condensation**
Co-condensation of the core-shell LP-MSN samples was achieved as follows: Solution 1: 100 mg $NH_4F$ (2.7 mmol), 21.7 mL $H_2O$ (1.12 mol), 2.97 ml TiPB (12 mmol) and 2.41 mL of a 25% CTAC solution (1.83 mmol) were mixed in a polypropylene reactor and heated to 60 °C under stirring. A second solution was prepared containing 12.77 mL TEA ( 97 mmol) to which was added 1.96 mL TEOS ( 8.8 mmol) combined with 0.2 mL APTES (0.86 mmol). This solution was heated to 90 °C under static conditions for 1 h and was subsequently added under strong stirring to solution 1. The combined solutions were allowed to cool to room temperature under stirring. After 20 minutes we added 0.1 mL TEOS ( 0.45 mmol) dropwise and stirred the solution for another 30 minutes. After this time the ingredients for the shell layer were added, consisting of a premixed solution of 22 µL MTES (0.11 mmol) and 22 µL TEOS (0.10 mmol). The condensation reaction was allowed to continue over night.

Subsequently, this solution was mixed with an additional 50 ml ethanol for 15 minutes, and the sample was collected by centrifugation. Template extraction followed immediately, as described for the LP-MSN samples.

The silane concentrations used in samples MP-1 to MP3 relate to the following mol% of silane coupling agents with respect to the total amount of silanes used:
MP1: core 0.2 mL APTES (0.86 mmol, or 9 mol%), shell 22 µL MTES (0.11 mmol or 1 mol%)
MP2: core 0.2 mL APTES (0.86 mmol, or 9 mol%), shell 22 µL MTES (0.11 mmol or 1 mol%)
MP3: core 0.1 mL APTES (0.43 mmol, or 4.3 mol%), shell 44 µL MTES (0.22 mmol or 2 mol%)

In order to achieve larger pores in sample MP-1, we centrifuged the reaction mixture after stirring over night at room temperature, exchanged the supernatant against 50 mL of a 1:1 $H_2O$/ethanol solution and continued stirring for 2 more days at room temperature. Template extraction was subsequentially performed as described above. Sample composition was estimated from TGA measurements, and the sulfur content was confirmed by EDX.

**Synthesis of MP-MSN-S-S:**

The amino groups present in the parent MSN samples were transformed first into carboxy groups by dissolving 100 mg succinic anhydride in 8 mL dry DMF and adding 20 mg of the MSN sample to the solution, which was stirred for 3 hour at room temperature. Samples were washed 3 times in ethanol and stored in ethanol (MSN-COOH).
To 5 mg of MSN-COOH suspended in 5 ml ethanol were added 5 µL EDC and 5 mg NHS-sulfo, and the mixture was stirred for 1 hour. This was followed by the dropwise addition of 0.5 mL of an ethanolic solution containing 24 mg cystamine dihydrochloride. This solution was stirred over night, and the MSN sample was washed 3 times by centrifugation and resuspension in ethanol, resulting in samples MP-MSN-S-S.

**siRNA loading and desorption**
siRNA concentrations were determined by UV measurements performed with the Nanodrop 2000c spectrometer (Thermo Scientific), with the nucleic acid module (sample volume 1.5 µL). siRNA adsorption was performed with aliquots of MSN samples, usually amounts of 100 µg that were exposed to 100 µL siRNA solutions (either in water or MES buffer solution at pH = 5) of predetermined concentration. Samples were vortexed and shaken at 37 °C for defined adsorption times between 15 minutes to several hours. Subsequently, samples were centrifuged (14000 rpm, 7 minutes) and the supernatant was measured again with the Nanodrop to determine the adsorbed amount by difference calculations.
To study the desorption process, the supernatant from the loading process was taken off by micropipette and was replaced with 100 µL PBS buffer desorption solution at pH = 7.4. The cumulative desorption was measured in the supernatant solution after centrifugation at preset time intervals. Samples were vortexed and again shaken after each measurement without change of the buffer solution.

**Polymer attachment**
**a) block copolymer (oligomer)**
The oligomer 454 was attached to MP-MSN samples after loading with siRNA. The oligomer was added directly to the loading solution after the complete siRNA uptake had been confirmed by Nanodrop analysis and after redispersion of the sample. Usually, 50 µg of oligomer 454 were added to 100 µg MSN and shaken for 1 h at 37 °C. This was followed by a 7 minute centrifugation at 14000 rpm. The supernatant was taken off (and measured as a reference in the cell transfection experiments) and was replaced with PBS buffer at pH = 7.4. Cell transfection was performed shortly thereafter.

**b) DOTAP**
A DOTAP layer was attached to MP-MSN samples after loading with siRNA. Here, the supernatant loading solution was removed, and to 100 µg of MSN sample 25 µL of a 30 wt% DOTAP solution (2.5 mg/mL in 60:40 $H_2O$:EtOH) was added by micropipette and was carefully redispersed with the pipette tip, followed by short sonification for 2 seconds. 225 µL cold water (4°C) was subsequently added and again mixed with the pipette for 30 seconds. A 2-fold washing in 100 µl sterile PBS at pH = 7.4 (centrifugation for 3.5 minutes at 14000 rpm) (was performed to remove excess lipid. The final sample was kept in PBS buffer for cell transfection.

**Characterization**

**Nitrogen sorption measurements**
Nitrogen sorption measurements were performed on a Quantachrome Instruments NOVA 4000e. All samples (15-20 mg) were heated to 393 K overnight while being evacuated (10 mTorr) to remove any adsorbates before nitrogen sorption was measured at 77 K. The BET (Brunauer-Emmett-Teller) surface areas were calculated from the corresponding nitrogen sorption isotherms in the range $p/p_0$ = 0.05 – 0.2. Pore size distribution curves were obtained using the non-local density functional theory (NLDFT) by applying either the equilibrium model or the adsorption model for cylindrical pores with nitrogen on silica as the reference module. Pore volumes were determined in LP-MSN samples at a relative pressure $p/p_0$ = 0.98 and in MP-MSN at a relative pressure $p/p_0$ = 0.8 to avoid inclusion of the interparticle textural porosity that is usually observed with nanoparticle powders (visible as a second hysteresis loop; see MP-MSN in Fig. 2).

**Infrared spectroscopy**
FTIR measurements were performed on a ThermoScientific Nicolet iN10 (MX) IR microscope equipped with a MTC-A liquid nitrogen cooled detector. Samples (about 1 mg) were prepared from an ethanolic solution that was dropped onto a non-transparent reflective sample holder and dried before measurements. Spectra were taken in transflection mode using a gold disc as reference. Spectra are all baseline subtracted, cleared from the atmospheric $CO_2$ absorption and normalized to the silica-based mode at 1075 $cm^{-1}$.

**Raman spectroscopy**
Raman spectra were obtained with a combined FTIR/Raman instrument (Bruker Equinox 55 with FRA-106 Raman attachment), providing excitation with an Nd:YAG (YAG: yttrium–aluminum–garnet) laser at 1064 nm. Spectra were collected in back reflection with a laser power of 100 mW.

**Thermogravimetry**

TGA measurements were performed with the silane-grafted LP-MSN and the co-condensed MP-MSN samples (about 10 mg of dried powder) on a Netzsch STA 440 thermobalance (heating rate of 10 K/min in a stream of synthetic air of about 25 mL/min). The loading of functional groups was determined in comparison to the purely siliceous parent samples.

**Transmission electron microscopy (TEM)**

TEM measurements were performed on a Jeol JEM-2011 microscope operating at 200 kV with a CCD detection unit. Samples were dispersed in ethanol and one drop of the resulting solution was then dried on a carbon-coated copper grid.

**Zeta potential measurements**

Zeta potential measurements were performed on a Malvern Zetasizer-Nano instrument equipped with a 4 mW He-Ne-Laser (λ=633 nm) and an avalanche photodiode detector coupled to a zetasizer titration system (MPT-2). pH titration was performed with about 0.5 mg of samples diluted in 10 mL bi-distilled water, with diluted NaOH and HCl solutions serving as titrants.

**Cell Microscopy**

For imaging, cells were seeded into 8-well ibiTreat slides (ibidi) at densities of 5000-10 000 cells per well the day prior to particle incubation. 3.5 µg of particles were added per well (300 µL volume) and removed after an incubation time of 45 minutes at 37°C by exchanging the medium. Imaging was performed 22 h and 48 h after incubation at 37 °C under a 5% $CO_2$ humidified atmosphere on live cells using spinning disc microscopy (Zeiss Cell Observer SD utilizing a Yokogawa spinning disk unit CSU-X1). The objective was a 1.40 NA 63x Plan apochromat oil immersion objective (Zeiss). Cy5 was imaged with 639 nm and Atto 488 with 488 nm laser excitation, respectively. For two color detection a dichroic mirror (560 nm, Semrock) and band-pass filters 525/50 and 690/60 (both Semrock) were used in the detection path. Separate images for each fluorescence channel were acquired using two separate electron multiplier charge coupled device (EMCCD) cameras (PhotometricsEvolve[TM]). Cell membranes were stained with wheat germ agglutinin Alexa Fluor 488 conjugate at a final concentration of 5 µg/ml.

**siRNA binding assay**

A 2.5% agarose gel was prepared by dissolving agarose in TBE buffer (Trizmabase 10.8 g, boric acid 5.5 g, disodium EDTA 0.75 g, and 1 L of water) under heating. After cooling down to about 50 °C and addition of GelRed™ (1:10 000) the agarose gel was cast into the electrophoresis unit. Samples containing 500 ng siRNA were prepared as described above and placed into the sample pockets after 4 µL of loading buffer (prepared from 6 mL of glycerol, 1.2 mL of 0.5 M EDTA, 8 mL of $H_2O$, 0.02 g of bromophenol blue) had been added. Electrophoresis was performed at 120 V for 40 min if not stated otherwise.

**Cell culture**

Human KB/eGFPLuc cells stably expressing luciferase (eGFP-luciferase fusion gene under the control of the CMV promoter) were cultivated in folate-free RPMI-1640 medium, supplemented with 10% fetal bovine serum (FBS), 100 U/mL penicillin and 100 µg/mL streptomycin.

## Gene silencing with siRNA

Gene silencing experiments were performed in KB/eGFPLuc cells. The siRNAs employed here were either siRNA against eGFP for silencing the eGFPLuc fusion protein or its negative control sequence siCtrl. As internal standard (100% control) we used cells that were treated only with a buffer solution (HBG: 20 mM HEPES buffered, 5% glucose, pH 7.4). Silencing experiments were performed in triplicates in 96-well plates. 24 h prior to transfection plates were coated with collagen and 4000 cells/well were seeded. Before transfection, the medium was replaced with 80 μL fresh growth medium. 20 μl of MSN suspension (containing usually 10 μg MSN, but also as low as 0.06 μg MSN in PBS buffer at pH = 7.4, prepared as described above) were added to each well and incubated at 37 °C. The medium was replaced after the indicated incubation time. 48 h after initial transfection, cells were treated with 100 μL cell lysis buffer per well. Luciferase activity in 35 μl cell lysate was measured using a Centro LB 960 plate reader luminometer (Berthold Technologies, Bad Wildbad, Germany) and a luciferin-LAR (1 M glycylglycine, 100 mM $MgCl_2$, 500 mM EDTA, DTT, ATP, coenzyme A) buffer solution. The relative light units (RLU) were presented as percentage of the luciferase gene expression obtained with buffer treated control cells.

## S1: Adsorption-desorption studies of siRNA in LP-MSN

LP-MSN were prepared with APTES and PhTMS silane coupling reagents to result in the following concentrations of functional groups throughout the particle body:
LP-2 with ca. 9 mol% $NH_2$, LP-3 with ca. 4 mol% Ph and 6 mol% $NH_2$, LP-3 with ca. 6 mol% Ph and 3 mol% $NH_2$. 1 mg of these samples was exposed to three successive aliquots of a loading solution containing each 16 μg siRNA. The loaded samples were centrifuged and first redispersed in water, resulting in nearly no RNA desorption within 3 hours. A medium exchange for PBS showed a cumulative release up to about 5 % of the adsorbed siRNA, while larger elution amounts were only observed after additional medium exchanges after 3 and 24 h increasing the total RNA release to about 25% after 4 days.

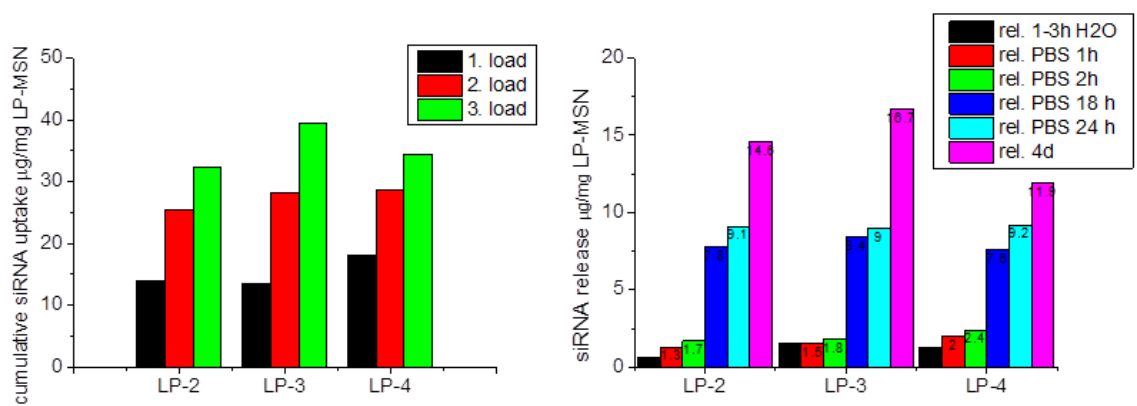

**S1:** cumulative siRNA adsorption and associated release diagrams of sample LP-2, LP-3 and LP-4

## S2: Transforming amino groups into disulfide-coupled amino groups

100 mg succinic anhydride, dissolved in 8 mL dry DMF was administered under nitrogen atmosphere to an ethanolic solution of 20 mg of sample S1 and was reacted at room temperature under stirring for 3 hours. The transformation of the amino groups into carboxy groups results in a shift of the IEP from about 6 to 4.6 as shown in Figure S2a). Additionally, carboxy groups are seen in the FTIR-spectrum by a C=O stretching frequency at 1714 cm$^{-1}$. This sample was subsequently conjugated with cystamine dihydrochloride in the following way: 5 µL EDC and 5 mg NHS-sulfo were added to a solution of 5 mg of the sample in 5 ml ethanol and stirred for 1 hour. An ethanolic solution of 24 mg cystamine was added dropwise under stirring and reacted overnight. The sample was retrieved by centrifugation and washed with ethanol 3 times. The conversion into now amino-terminated cleavable residues is seen by a return of the IEP to a higher value of 5.9 in the zeta potential curve. Furthermore, the C=O stretching bond has vanished and typical amide stretching frequencies of 1655 and 1542 cm$^{-1}$ are seen for the amide I and amide II bonds in the FTIR.

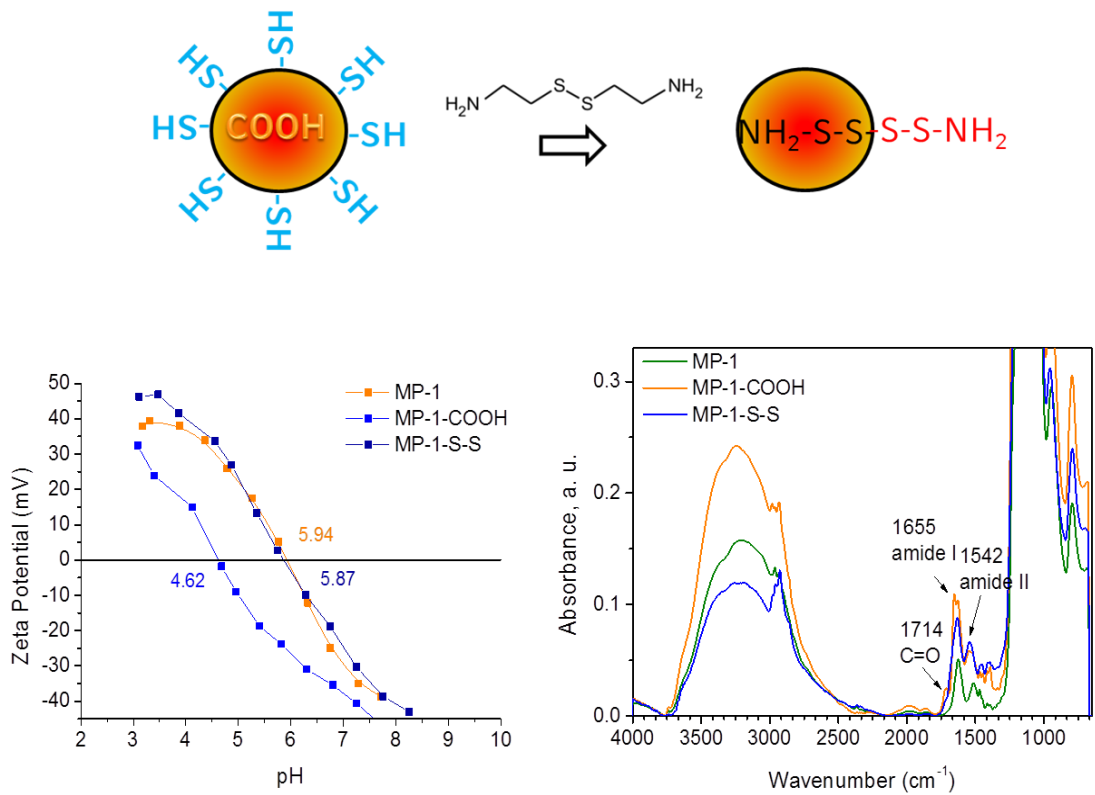

**S2:** Transforming covalently coupled amino groups into disulfide-coupled amino groups: left) zeta-potential measurement of the parent sample MP-1, after carboxylation and disulfide coupling of amino groups via csytamin addition; right) the corresponding FTIR measurements

**S3: siRNA in sample MP-3 with lower amino group concentration (core 4 mol% NH$_2$, shell 2 mol% SH)**

Adsorption was performed in MES buffer at pH=5 and the release was performed with PBS buffer at pH=7.4. The siRNA loading process is substantially slowed down compared to samples MP-1 and MP-2 with 9% amino groups, however, 93% of the offered RNA was absorbed after 5 h amounting to 72 µg/mg with the highest tested siRNA loading solution of 80 µg/mL. The diminished charge in this sample led to an easier elution of RNA in PBS buffer. Now, 60-70% were immediately released (measured after 15 minutes) and 68 to 87% of free RNA were measured after 1 day.

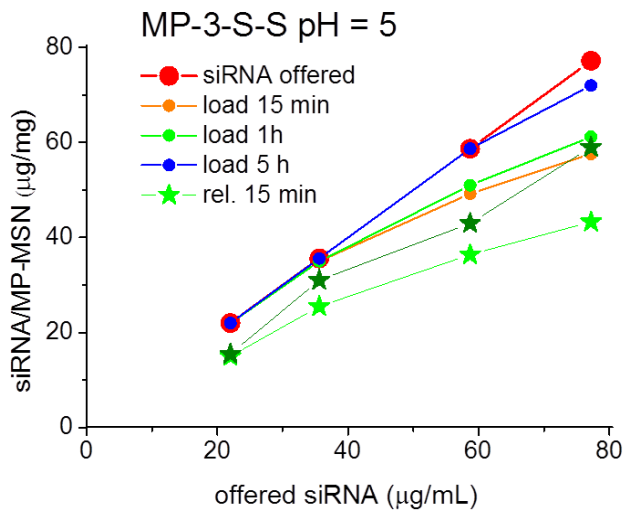

**S3a:** Cumulative siRNA sorption and release curves of sample MP-3-S-S containing 4 mol% NH$_2$ in the core and 2 mol% SH in the shell, without polymer capping.

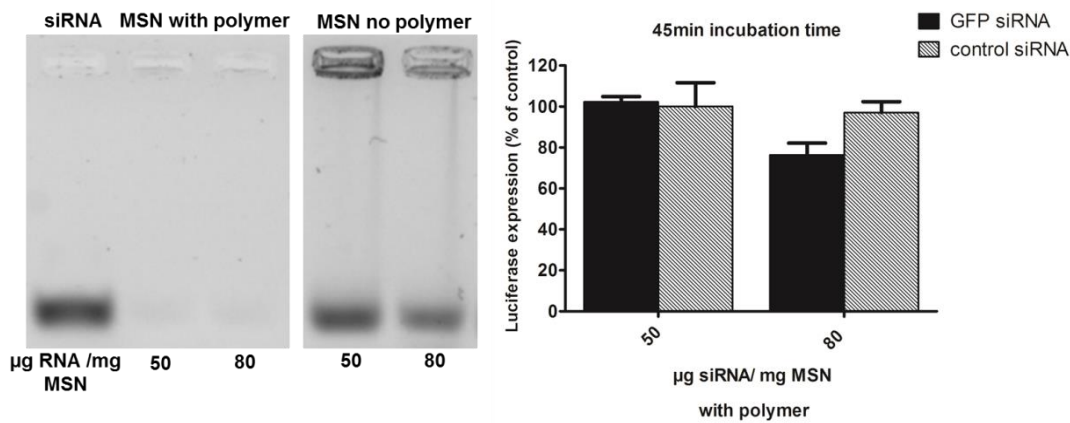

**S3b:** Gel electrophoresis and transfection of RNA-loaded sample MP-S3-S-S

Sample MP-S3-S-S with 4 mol% NH$_2$ and 2 mol% SH is loaded with 50 and 80 µg siRNA/mg MSN. Gel electrophoresis is shown with and without being capped with the block-copolymer. SiRNA is still present in the carrier system in the uncapped state, while the electrostatic interaction with the cationic polymer extracted the siRNA from the carrier and was detected in the supernatant solution. Luciferase knockdown is thus only experienced to a very small extent in the higher loaded sample in the experiments.

## S4: Spectroscopic evidence for mercapto groups and their transformation into disulfide-coupled amino groups

The mercapto S-H stretching frequency is seen at 2576 cm$^{-1}$ in the Raman spectrum of the parent samples MP-1 and MP-3, however, it is better visible in sample MP-3 due to a higher concentration of 2 mol% mercapto groups in the shell. After reaction with cystamine this indicative band has completely vanished. Consequently, the IEP of the parent sample MP-3 shifts from 4.8 to 6.2 after cystamine binding.

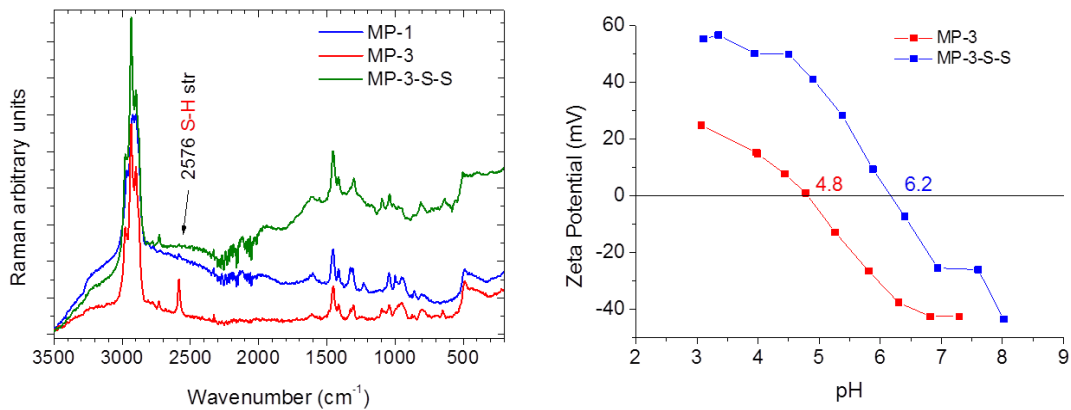

**S4**: Raman spectroscopy of sample MP-1 and MP-3 and MP-3 after reaction with cystamine and zeta potential measurements of MP-3 before and after reaction with cystamine

## S5: Evidence for polymer attachment to siRNA loaded MSNs

The IEP of the parent compound MP-1 (purple) shifts from 5.9 after loading with siRNA (blue) to 5.1 and returns to 6.4 after attachment of the cationic polymer (green); the zeta potential of the pure polymer is shown in addition (orange; IEP 7.6)

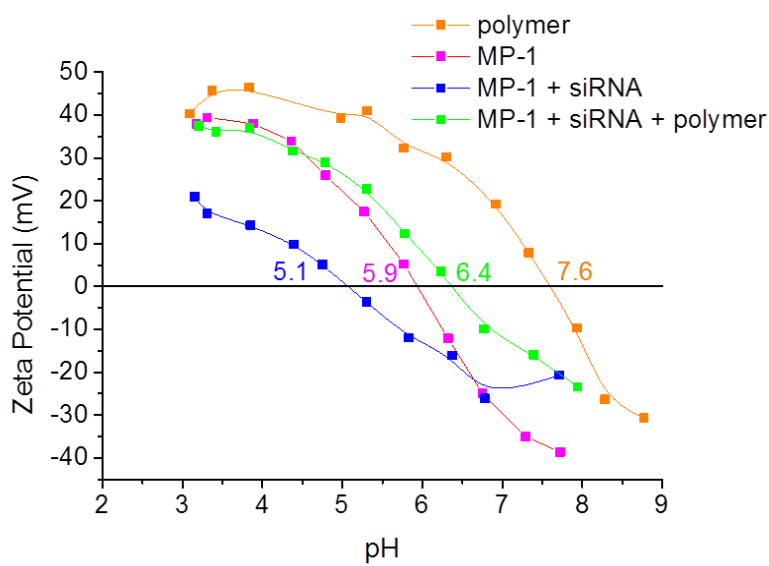

**S5:** Zeta potential measurements of sample MP-1 before and after RNA loading and polymer capping

**S6: Knock-down efficiencies of block-copolymer 454 capped MSNs with different pore-sizes**

SiRNA loadings were varied from 5 wt%, over 10 wt% to 13 wt% siRNA, while keeping the total amount of siRNA at 0.5 µg/well constant (less particles /well). A sample without block copolymer attachment is also shown. An incubation time of 45 minutes was used, followed by a read out after 48 h. These results are very similar to those obtained with sample MP-1 under the same conditions as shown in Fig. 7 in the main text, which are added here again for easier comparison.

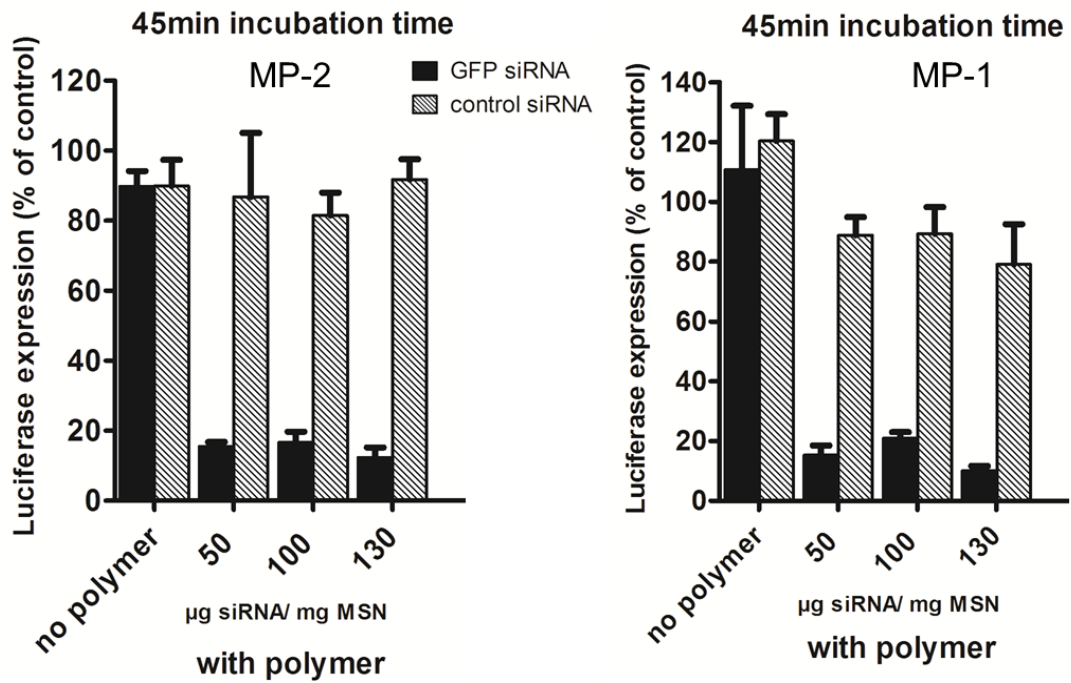

**S6:** Cell transfections with sample MP-2 with 3.9 nm pores in comparison to MP-1 with 4.7 nm pores

**S7: Knock-down efficiencies of DOTAP-capped MSNs**

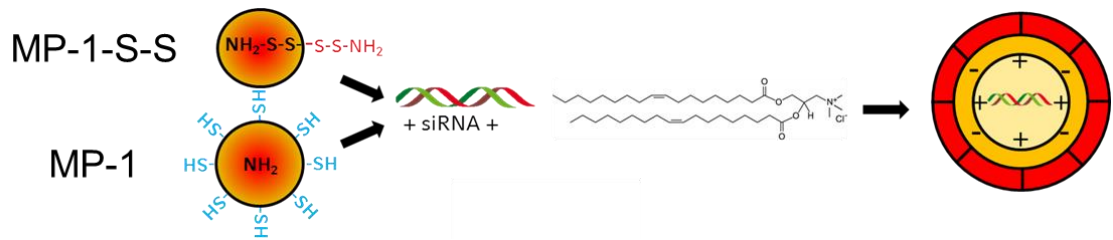

**S7:** scheme for sample assembly of MP-1 and MP-1-S-S covered with a **DOTAP** cationic lipid layer

DOTAP was administered to MP-1 and MP-1-S-S subsequently to siRNA loading. Samples were centrifuged and the supernatant containing surplus DOTAP was removed and exchanged for a PBS buffer solutions. These samples were used for cell transfection experiments shown below. Cells were incubated for 4 h before the medium was exchanged. Read-out was performed after 48 (7a, 7b) or 72 h (7a).

Samples were additionally treated with chloroquine or the endosomolytic peptide INF-7 for comparison. The Luciferase activity was reduced to over 80% and neither chloroquine nor INF-7 was necessary to increase the efficacy. It is noted that the toxicity was increased upon INF-7 addition.

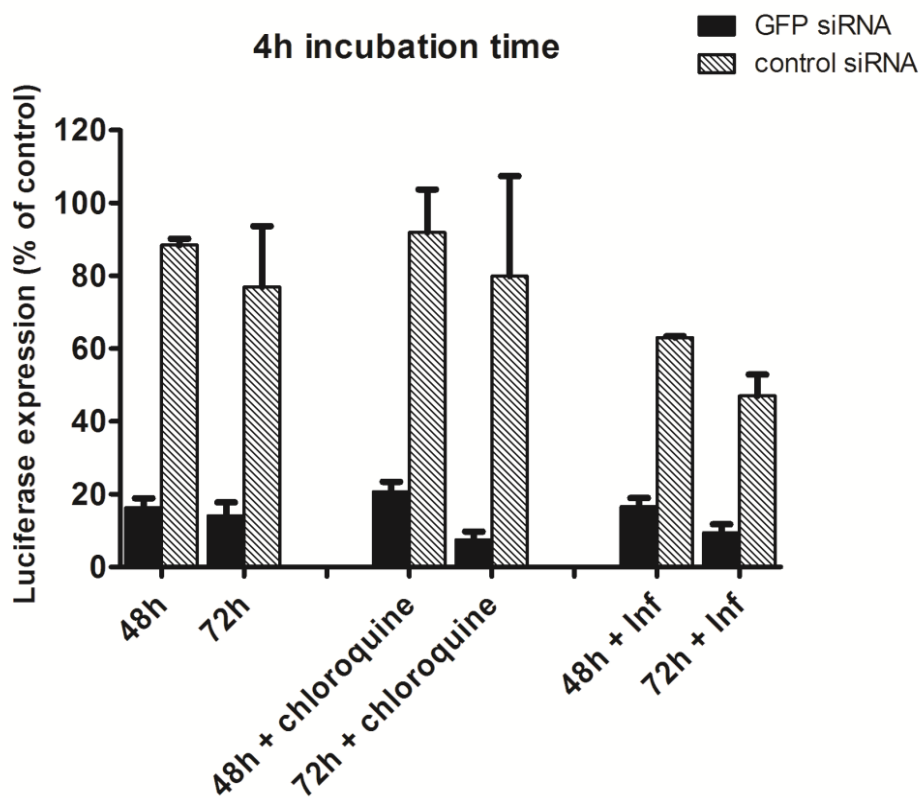

**S7a)** Transfection with DOTAP covered sample MP-S-S

SiRNA concentration effects were measured with sample MP-1 and MP-1-S-S in combination with DOTAP. The loading with siRNA was increase from 5 wt% to 10 wt% and even to 200 wt% while the total siRNA amount administered per well was kept at 0.5 µg/well.

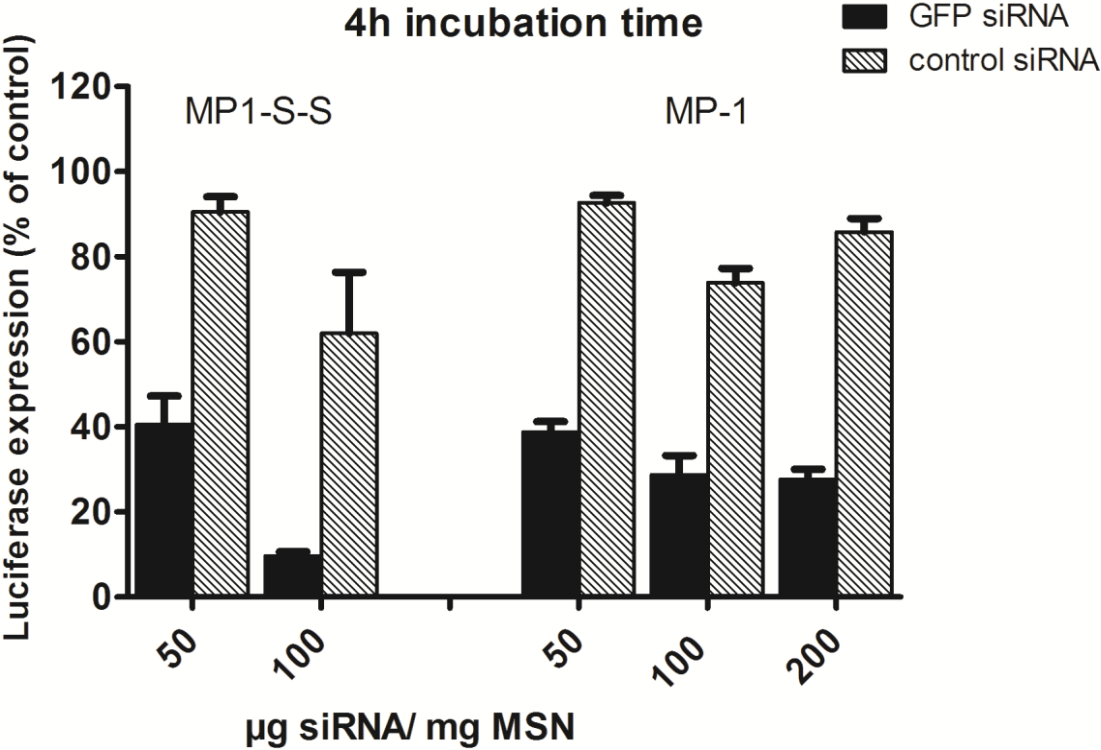

**S7b**) Transfection after 48 h with DOTAP covered samples MP-1-S-S in comparison to MP-1 with increasing siRNA loadings